\documentclass[journal]{IEEEtran}
\usepackage{graphicx}
\usepackage[cmex10]{amsmath}
\usepackage{cases}
\usepackage[tight,footnotesize]{subfigure}
\usepackage{amsthm}
\usepackage{cite}
\usepackage{amssymb}
\usepackage{algorithm}
\usepackage{algorithmic}
\usepackage{multirow}
\usepackage{amsmath}
\usepackage{xcolor}
\usepackage{CJK}
\usepackage{subeqnarray}
\usepackage{cases}
\usepackage{enumerate}
\usepackage{cuted}
\usepackage{booktabs}
\usepackage{mathrsfs}
\usepackage{bbm}
\usepackage{array}
\usepackage[colorlinks]{hyperref}
\newcolumntype{M}[1]{>{\centering\arraybackslash}m{#1}}

\newtheorem{theorem}{\hskip\parindent \it {Theorem}}
\newtheorem{remark}{\hskip\parindent\it{Remark}}

\ifCLASSINFOpdf
\else
\fi

\begin{document}
\title{Frequency Diverse (FD)-RIS-Enhanced Covert Communications: Defense Against Wiretapping  via Joint Distance-Angle Beamforming}
\author{Han Xiao,~\IEEEmembership{Graduate Student Member,~IEEE,} Xiaoyan Hu$^*$,~\IEEEmembership{Member,~IEEE,} Wenjie Wang,~\IEEEmembership{Senior Member,~IEEE,} \\
Kai-Kit~Wong,~\IEEEmembership{Fellow,~IEEE}, Kun~Yang,~\IEEEmembership{Fellow,~IEEE}, Chan-Byoung Chae,~\IEEEmembership{Fellow,~IEEE}
	
\thanks{H. Xiao, X. Hu, and W. Wang are with the School of Information and Communications Engineering, Xi'an Jiaotong University, Xi'an 710049, China. (email: hanxiaonuli@stu.xjtu.edu.cn, xiaoyanhu@xjtu.edu.cn, wjwang@mail.xjtu.edu.cn. \emph{(Corresponding author: Xiaoyan Hu.)}}
\thanks{K. K. Wong is with the Department of Electronic and Electrical Engineering, University College London, Torrington Place, WC1E 7JE, U.K., and he is also with Yonsei Frontier Lab, Yonsei University, Seoul, Korea. (email: kai-kit.wong@ucl.ac.uk)}
\thanks{K. Yang is with the State Key Laboratory of Novel Software Technology, Nanjing University, Nanjing 210008, China, and School of Intelligent Software and Engineering, Nanjing University (Suzhou Campus), Suzhou 215163, China. (email: kunyang@nju.edu.cn).}
\thanks{C.-B. Chae is with the School of Integrated Technology, Yonsei University, Seoul, 03722, Korea (email: cbchae@yonsei.ac.kr).}
}

\maketitle

\begin{abstract}
In response to the ``\textit{security blind zone}'' challenges faced by traditional reconfigurable intelligent surface (RIS)-aided covert communication (CC) systems, the joint distance-angle beamforming capability of frequency diverse RIS (FD-RIS) shows significant potential for addressing these limitations. Therefore, this paper initially incorporates the FD-RIS into the CC systems and proposes the corresponding CC transmission scheme. Specifically, we first develop the signal processing model of the FD-RIS, which considers effective control of harmonic signals by leveraging the time-delay techniques. The joint distance-angle beamforming capability is then validated through its normalized beampattern. Based on this model, we then construct an FD-RIS-assisted CC system under a multi-warden scenario and derive an approximate closed-form expression for the covert constraints by considering the worst-case eavesdropping conditions and utilizing the logarithmic moment-generating function. An optimization problem is formulated which aims at maximizing the covert user's achievable rate under covert constrains by jointly designing the time delays and modulation frequencies. To tackle this non-convex problem, an iterative algorithm with assured convergence is proposed to effectively solve the time-delay and modulation frequency variables. To evaluate the performance of the proposed scheme, we consider three communication scenarios with varying spatial correlations between the covert user and wardens. Simulation results demonstrate that FD-RIS can significantly improve covert performance, particularly in angular-overlap scenarios where traditional RIS experiences severe degradation. These findings further highlight the effectiveness of FD-RIS in enhancing CC robustness under challenging spatial environments.
\end{abstract}
\IEEEpeerreviewmaketitle

\section{Introduction}\label{sec:S1}
\IEEEPARstart{W}{ith} the rapid development and deployment of fifth-generation (5G) as well as the forthcoming sixth-generation (6G) communication technologies, the vision of a fully interconnected world enabled by the Internet of Things (IoT) is becoming a reality. Consequently, the number of connected devices is growing at an unprecedented rate, generating massive volumes of real-time data. These data often includes highly sensitive information such as financial transactions and medical records, and are typically transmitted over open wireless channels that are inherently vulnerable to eavesdropping \cite{zheng2019multi}. Under such circumstances, the challenges related to communication security and privacy are becoming increasingly critical. While cryptographic techniques \cite{peng2020covert} and physical layer security (PLS) \cite{liang2009information, shiu2011physical} methods offer protections at the content level, they often fail to prevent adversaries from detecting the existence of communications,  which may leave security risks for unauthorized users to launch attacks. Recently, the technology of covert communication (CC) has emerged as a novel security paradigm that aims to conceal the existence of communications, thereby significantly reducing the risks of being detected by adversaries and  providing a higher level of security than cryptographic and PLS techniques \cite{xiao2024star, yan2019low, li2025covert}.

\subsection{Related Works}
As a groundbreaking work, Bash \textit{et al.} \cite{bash2013limits} initially demonstrated that, under the assumption of additive white Gaussian noise (AWGN) channel between transmitter and receiver/warden, each subject to non-zero noise power, it is possible to reliably and covertly transmit $\mathcal{O}(\sqrt{n})$ bits from transmitter to receiver over $n$ channel uses. Actually, this conclusion may be overly pessimistic, as the analysis in \cite{bash2013limits} does not fully consider the inherent uncertainties that exist in practical communication environments, particularly those arising from background noise and wireless channel. Overlooking these factors can result in an overestimation of the adversary's detection capability. To this end, the authors in \cite{goeckel2015covert} and \cite{wang2018covert} concluded that $\mathcal{O}(n)$ bits of information can be reliably and covertly transmitted from the transmitter to receiver by taking into account the uncertainties of the warden's background noise power and the channel between the relay and the warden. To further improve covert performance, artificial uncertainties have been introduced into communication systems to confuse potential wardens \cite{li2020optimal, xiong2020covert, hu2019covert}. 
Specifically, \cite{li2020optimal, xiong2020covert} proposed the use of external jamming nodes to generate random interference, thereby introducing additional uncertainty and impairing the warden’s detection ability. These studies also provided a systematic analysis of the covert performance gains achieved through such jamming-based uncertainty. 
Furthermore, \cite{hu2019covert} employed full-duplex receivers that simultaneously receive covert signals and transmit artificial noise to degrade the warden's detection performance.

While the aforementioned methods have demonstrated notable effectiveness in enhancing the covertness of communication systems, their performance heavily depends on the wireless propagation environment, particularly in high-frequency systems. This strong environmental dependence limits their practicality and constrains the achievable covert performance gains. To address this issue, the technology of reconfigurable intelligent surface (RIS)  has emerged as a promising solution \cite{huang2019reconfigurable, zhu2025transmissiveris}, owing to its ability to manipulate the electromagnetic properties (phase shifts and amplitudes) of incident signals, thereby establishing a controllable end-to-end virtual channel between base stations (BS) and users. As a result, RIS has been widely integrated into various communication systems such as mobile edge computing (MEC) \cite{xiao2025STAR-RIS_UAV, xiao2025Energy-Efficient, hu2021reconfigurable}, integrate sensing and communications (ISAC) \cite{yu2024activeRIS, zhu2023joint}, and CC networks \cite{lu2020intelligent, wang2021intelligent, lv2021covert, hu2025reconfigurable}.
Specifically, Lu \textit{et al.} \cite{lu2020intelligent} investigated the potential of RIS to enhance CC performance under the uncertainty of the warden's background noise power. Simulation results showed significant performance improvement compared to systems without RIS. Building on this, Wang \textit{et al.} \cite{wang2021intelligent} extended RIS into a multi-antenna CC system with a full-duplex receiver capable of receiving confidential signals while transmitting artificial noise, where the RIS not only assists covert transmission but also smartly controls the interference to further impair the eavesdropping accuracy.
Furthermore, Lv \textit{et al.} \cite{lv2021covert} explored RIS-assisted CC in a NOMA-based uplink and downlink system, analyzing how RIS-induced phase shift uncertainty affects the warden's detection. This work offered new insights into the applicability of RIS in CC. More recently, Hu \textit{et al.} \cite{hu2025reconfigurable} investigated a more practical CC scenario involving multiple public users and one covert user, jointly optimizing active and passive beamforming to maximize the covert rate while ensuring quality of service (QoS) for public users.

\subsection{Motivation and Contributions}
In fact, although the strong spatial manipulation capability of RIS has significantly improved the covert performance of communication systems, existing RIS-assisted CC schemes\footnote{Note that unless otherwise specified, the system considered in this paper operates under the far-field communication model.} still suffer from the following limitations:
(\romannumeral 1) Most of the existing studies assume that the warden and the legitimate users are spatially uncorrelated, meaning their channels are statistically independent. However, in practical scenarios, an intelligent warden may actively move within the service area to find a more favorable position for detection. Such dynamic behavior inevitably introduces spatial correlation between the warden and legitimate users. As a result, RIS-assisted CC schemes often exhibit poor adaptability in scenarios with high spatial correlation, thereby limiting their practical applicability and robustness.
(\romannumeral 2) Due to the planar wave propagation characteristics of electromagnetic signals in the far field, current RIS-assisted far-field CC systems still suffer from the issue of ``\textit{security blind zone}”. Under the plane wave model, RIS can only focus or enhance signals in specific directions but lacks the ability to distinguish targets at different distances \cite{liu2024ris}. In other words, when the warden locates close to or overlaps with the legitimate user in angle, RIS cannot distinguish between them in terms of distance. These limitations make it difficult to guarantee the CCs between the BS and legitimate user.

To address the aforementioned challenges, the technology of frequency diverse RIS (FD-RIS)  \cite{xiao2024frequency} offers a feasible and effective solution. Its core principle involves applying time modulation to periodically vary the phase shifts of RIS's elements. Based on Fourier series theory, this modulation scatters the incident signal into a series of harmonic components, each with a frequency offset relative to the carrier, determined by the harmonic order and modulation period. Unlike conventional RIS, FD-RIS introduces these frequency offsets to enable fine-grained control of signal propagation in both distance and angle domains. This dual-domain beamforming provides greater flexibility and higher control accuracy at the physical layer, thereby offering significant potentials to overcome the limitations of the traditional RIS-assisted CC systems.
While the theoretical model and joint distance-angle beamforming capability of FD-RIS have been systematically analyzed and validated in \cite{xiao2024frequency}, its application in practical communication scenarios remains under-explored. To further advance FD-RIS and tackle the challenges associated with the traditional RIS-assisted CC systems, this paper, for the first time, integrates FD-RIS into CC systems and proposes a corresponding covert transmission scheme.
The main contributions of this paper are summarized as follows:
\begin{itemize}
\item \textit{\textbf{Signal Processing Model of FD-RIS with Pure Harmonic Signals and Time delays:}} The signal processing model of the FD-RIS is first established by considering the implementation of time modulation at each element. Specifically, to enable effective control over the reflected harmonic signals, a novel time modulation scheme is employed to periodically adjust the phase shifts of each RIS element. With this method, a desired single-order harmonic component can be selectively extracted, while all other harmonic components are effectively suppressed. To overcome the limitation on spatial degrees of freedom (DoFs) imposed by the time modulation policy, additional time delays are incorporated into the modulation process. Based on the established signal processing model, we further derive the received signal expression at a given location and demonstrate the distance-angle beamforming capability of the FD-RIS through its corresponding beampattern analysis.
\item \textit{\textbf{Closed-form Expression Derivation for CC System Indicators:}} To ensure the reliability and applicability of the proposed FD-RIS-assisted CC scheme, an extreme scenario is considered in which the wardens are assumed to have access to various system parameters. Based on this assumption, from the wardens' perspective and considering uncertainty in their background noise power, we derive the closed-form expression for the optimal detection error probability. Furthermore, considering the BS Alice's uncertainty for the non-line-of-sight (NLoS) channel components between the FD-RIS and wardens, we further derive an approximate closed-form expression for the covert constraint from Alice's perspective using the logarithmic moment-generating function.
\item \textit{\textbf{Problem Formulation and Proposed Iterative Algorithm with Guaranteed Convergence:}} To validate the effectiveness of the FD-RIS in enhancing the covert performance of communication systems, an optimization problem is formulated to maximize the covert user's achievable rate under covert constraints, by jointly designing the time delays and modulation frequencies. To efficiently solve this non-convex problem, an alternative optimization strategy is adopted, which decomposes the original problem into two subproblems: a time-delay subproblem and a modulation frequency subproblem.
For the time-delay subproblem, the minimum mean-square error (MMSE) method combined with the penalty dual decomposition (PDD) framework is employed. For the modulation frequency subproblem, the successive convex approximation (SCA) technique is applied to approximate the non-convex problem into a convex form.
\item \textit{\textbf{Significant Performance Gain:}} To evaluate the potentials of FD-RIS in CC systems, three communication scenarios with varying levels of spatial correlation between covert user and the wardens are considered. In particular, the third scenario represents an extreme case where one of wardens is perfectly aligned in same angle with the covert user, posing a critical challenge for conventional RIS-assisted CC systems. Extensive numerical simulations are conducted, and the results show that the FD-RIS, leveraging its joint distance-angle beamforming capability, can significantly enhance covert performance even with angular overlap between the warden and legitimate user. In contrast, the performance gains of conventional RIS scheme degrade rapidly as the spatial correlation between wardens and covert user increases, with the most severe degradation observed in the third scenario.
\end{itemize}


\textit{Notation:} $\mathbb{Z}$ denotes the integer set. The operations $(\cdot)^T$, $(\cdot)^*$  and $(\cdot)^H$ denote the transpose, conjugate and conjugate transpose, correspondingly.  $\operatorname{Diag}(\mathbf{a})$ denotes a diagonal matrix whose diagonal elements are composed of the vector $\mathbf{a}$. Additionally, the symbols $|\cdot|$  and $\|\cdot\|_2$ are indicative of the complex modulus and spectral norm, respectively. Operator $\circ$ denotes the Hadamard product. $\mathcal{I}_i(\mathbf{a})$ represents the $i$-th entry of the vector $\mathbf{a}$. $\chi_1^2(a)$ and $\chi_2^2(a)$ represent non-central chi-square distributions with $1$ and $2$ degrees of freedom, respectively, where $a$ is the non-centrality parameter. $\operatorname{mod}(a, b)$ represents the operator of taking the remainder with $a$ and $b$ being the dividend and divisor, respectively.
\section{Signal Processing Model for FD-RIS}\label{sec:S2}
\begin{figure}[ht]
	\centering
\includegraphics[scale=0.4]{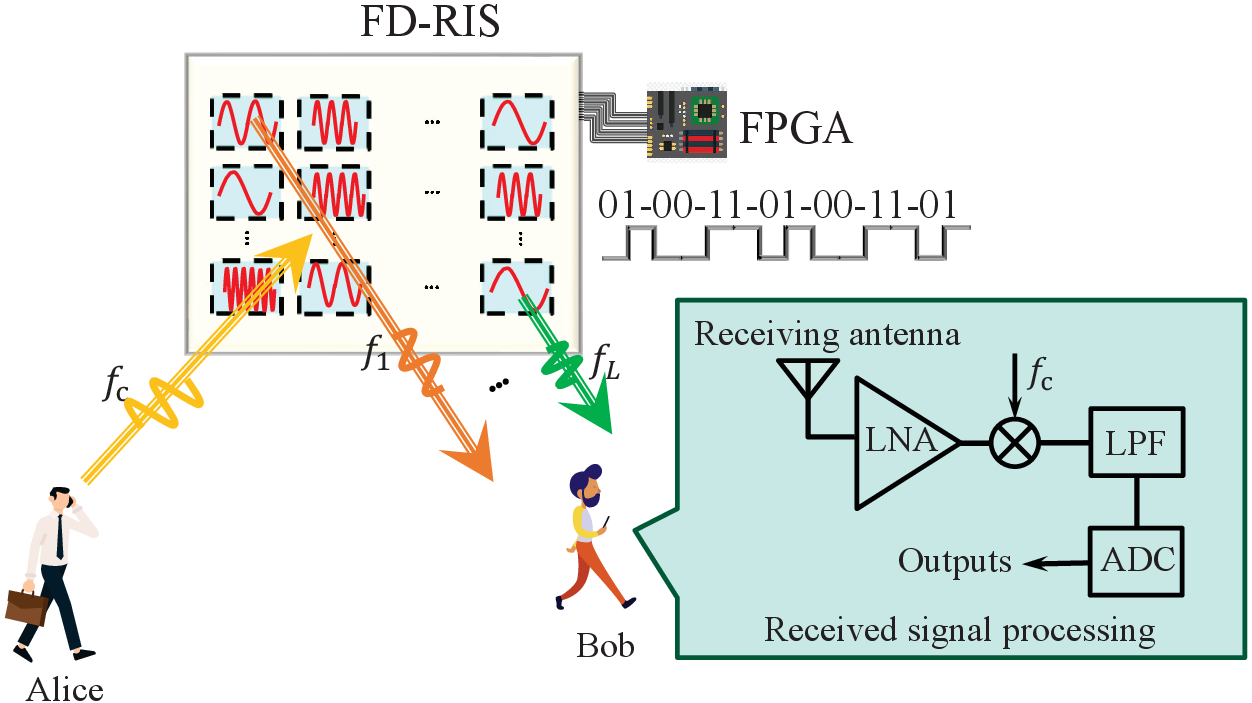}\\
	\caption{Signal transmission schematic diagram assisted by the FD-RIS.}\label{fig:FD-RIS_principle}
\end{figure}

In this section, we will derive the signal processing model of the FD-RIS \textcolor{blue}{based on the schematic diagram in Fig. \ref{fig:FD-RIS_principle}, which includes a transmitter Alice, a receiver Bob and  a FD-RIS with $L=L_yL_z$ elements which are uniformly deployed in the available space with distance $d$, where $L_y$ and $L_z$ are the number of elements along $y$ and $z$ axis, respectively.} Additionally, we assume that each element uses an independent time-modulation frequency to adjust its reflection coefficient under control of field programmable gate array (FPGA). In other words, the modulation frequency $\Delta f_l$ for $l=(l_z-1)L_y +l_y\in\mathcal{L}\triangleq\{1, \cdots, L\}$, is leveraged to modulate the reflection coefficient of the $l$-th element. Thus, when Alice with single antenna transmits the signal $x(t)=e^{j2\pi f_\mathrm{c}t}s(t)$, where $f_\mathrm{c}$ and $s(t)$ represent the carrier frequency and the narrow band complex envelope, to the FD-RIS, the reflected signal can be expressed as
\begin{align}
	x_\mathrm{r}(t)=\rho(d^\mathrm{ar})s(t)\sum_{l_y=1}^{L_y}\sum_{l_z=1}^{L_z}e^{j2\pi f_\mathrm{c}\left(t-\frac{d^\mathrm{ar}_{l_y, l_z}}{c}\right)}U_{l_y, l_z}(t),
\end{align}
where $\rho(d^\mathrm{ar})$ is the large-scale path loss associated with the spatial distance between Alice and the FD-RIS, i.e., $d^\mathrm{ar}$. $U_{l_y, l_z}(t)$ represents the periodic time-modulation reflected coefficient of the $(l_y, l_z)$-th element, which is a function of time-modulation frequency $\Delta f_l$. $c$ is the light speed. In addition, $d^\mathrm{ar}_{l_y, l_z}$ denotes the distance between Alice's antenna and the $(l_y, l_z)$-th element of the FD-RIS, which can be expressed as $d^\mathrm{ar}_{l_y, l_z}=d^\mathrm{ar}+\Upsilon^\mathrm{ar}_{l_y, l_z},$
where $\Upsilon^\mathrm{ar}_{l_y, l_z}=(l_z-1)d\cos(\theta^\mathrm{ar})+(l_y-1)d\sin(\theta^\mathrm{ar})\cos(\phi^\mathrm{ar})$ with $\theta^\mathrm{ar}$ and $\phi^\mathrm{ar}$ being the azimuth and elevation angles of arrival (AoA) for signal transmitted from Alice to the FD-RIS, respectively.

According to the Fourier series expansion theory, the periodic reflection coefficient $U_{l_y, l_z}(t)$ can be expressed as
$
	U_{l_y, l_z}(t)=\sum_{z=-\infty}^{z=+\infty}a_{l_y, l_z}^{z}e^{j2\pi z\Delta f_lt},
$
where $z\in\mathbb{Z}$ is the order number of the harmonic, $a_{l_y, l_z}^{z}$ denotes the Fourier series coefficient, given by
$
	a_{l_y, l_z}^{z}=\Delta f_l\int_{0}^{T_l}U_{l_y, l_z}(t)e^{-j2\pi z\Delta f_l t}dt,
$
with $T_l=\frac{1}{\Delta f_l}$ denoting the time-modulation period.
Thus, the reflected signal can be re-written as
\vspace{-2mm}\begin{align}\label{eq_reflection_signal}
	x_\mathrm{r}(t)	=&\rho(d^\mathrm{ar})s(t)\sum_{l_y=1}^{L_y}\sum_{l_z=1}^{L_z}e^{-j2\pi f_\mathrm{c}\frac{d^\mathrm{ar}_{l_y, l_z}}{c}}\times\notag\\
	&\sum_{z=-\infty}^{z=+\infty}a_{l_y, l_z}^{z}e^{j2\pi (f_\mathrm{c}+z\Delta f_l)t}. 
\end{align}
\begin{remark}
	As shown in \eqref{eq_reflection_signal}, FD-RIS will reflect the incident signal as a series of harmonic signals with frequencies $f_\mathrm{c}+z\Delta f_l,~ z\in\mathbb{Z}, l\in\mathcal{L}$. Actually, the presence of numerous harmonic signals may pose significant challenges to communication systems since it is difficult to precisely regulate all harmonic signals, 
which may result in the leakage of certain harmonics into undesired directions. This unintended scattering may severely interfere with the normal operations of other communication systems, degrading their overall communication quality. Moreover, the high-order harmonics often exceed the designated spectral range of the target communication system. This may cause spectrum pollution, interfering with other wireless systems operating in certain frequency bands. To address these potential challenges, developing effective techniques to suppress or manage these harmonic signals is a critical issue that needs to be resolved.
\end{remark}

To address the aforementioned challenges, we employ the periodic time-modulation technique, as detailed in \cite{dai2020high}, to periodically modulate the reflection coefficient of each element. Consequently, the time-modulated reflection coefficient for the $(l_y, l_z)$-th element can be expressed as
\begin{align}\label{eq_reflection_coefficient}
	U_{l_y, l_z}(t)=A_0e^{\phi_0+S\operatorname{mod}(t, T_l)},
\end{align}
where $A_0$ and $\phi_0$ denote the initial amplitude and phase-shift of each element, respectively. $S$ represents the phase slope. Hence, we can derive the Fourier coefficient as
$
	a_{l_y, l_z}^{z}
	=\frac{jA_0e^{j\phi_0}\left(1-e^{j(ST_l-2z\pi)}\right)}{ST_l-2z\pi}.
$
According to the derived expression of the Fourier series coefficient, it is observed that when $ST_l=2g\pi,~g\in\mathbb{Z}$, $a_{l_y, l_z}^{z}$ can be further expressed as
\begin{align}
	a_{l_y, l_z}^{z}=\begin{cases}
		0,& z\neq g,\\
		A_0e^{j\phi_0},& z=g,
	\end{cases}
\end{align}
which indicates that we can obtain any pure single-order harmonic signal with index $g$ by appropriately selecting the phase slope $S$. In other words, the reflection signals of each element only include the $g$-th order harmonic signal when $ST_l=2g\pi$. Therefore, if each element utilizes the $g$-th harmonic signal to transmit information, the signal reflected by the FD-RIS can be further expressed as
\begin{align}
		x_\mathrm{r}(t)=&\rho(d^\mathrm{ar})s(t)\sum_{l_y=1}^{L_y}\sum_{l_z=1}^{L_z}e^{-j2\pi f_\mathrm{c}\frac{d^\mathrm{ar}_{l_y, l_z}}{c}}A_0e^{j\phi_0}\times\notag\\
		&e^{j2\pi (f_\mathrm{c}+g\Delta f_l)t}.
\end{align}

Although the aforementioned time-modulation method facilitates the reflection of the single-order harmonic signal, it significantly constrains the spatial modulation capabilities of the reflected harmonic signal. Specifically, each element will provide the same reflection coefficient, i.e., $A_0e^{j\phi_0+j2\pi g\Delta f_l t}$, for the reflected harmonic signal after applying this time-modulation method.
 To address this challenge, the time delay, $\kappa_l,~ l\in\mathcal{L}$, is introduced into each element's periodic reflection coefficient. According to the time-delay feature of the Fourier series, the reflection coefficient can be further derived as
\begin{align}
		U_{l_y, l_z}(t)=A_0e^{j(\phi_0+2\pi g\Delta f_l t-2\pi g\Delta f_l \kappa_l)},
\end{align}
which indicates that each element can offer an arbitrary phase-shift for the reflected harmonic signal by selecting the suitable time delay. As a result, applying time delays to each element of the FD-RIS will greatly enhance its spatial modulation ability for incident signals. Under this condition, the reflected signal of the FD-RIS can be  expressed as
\begin{align}
		x_\mathrm{r}(t)=&\rho(d^\mathrm{ar})s(t)\sum_{l_y=1}^{L_y}\sum_{l_z=1}^{L_z}e^{-j2\pi f_\mathrm{c}\frac{d^\mathrm{ar}_{l_y, l_z}}{c}}\times\notag\\
	&A_0e^{j(\phi_0-2\pi g\Delta f_l \kappa_l)}e^{j2\pi (f_\mathrm{c}+g\Delta f_l)t}.
\end{align}

Therefore, the received signal of Bob located at point $(d^\mathrm{rb}, \theta^\mathrm{rb}, \phi^\mathrm{rb})$ is given by
$
	y(t)=\rho(d^\mathrm{rb})x_\mathrm{r}\left(t-d_{ly,lz}^\mathrm{rb}/c\right)+n_\mathrm{b}(t),
$
where $n_\mathrm{b}(t)\sim\mathcal{CN}(0, \sigma_\mathrm{b}^2)$ is the AWGN with $\sigma_\mathrm{b}^2$ being the noise power. $d^\mathrm{rb}$ denotes the spatial distance between the FD-RIS and Bob, $\theta^\mathrm{rb}$ and $\phi^\mathrm{rb}$ are the azimuth and elevation angles of departure (AoD) for signals reflected from FD-RIS to Bob, respectively. $d^\mathrm{rb}_{l_y, l_z}$ is the distance between the $(l_y, l_z)$-th element and Bob, which is given by
$
	d^\mathrm{rb}_{l_y, l_z}=d^\mathrm{rb}+\Upsilon^\mathrm{rb}_{l_y, l_z},
$
with $\Upsilon^\mathrm{rb}_{l_y, l_z}=(l_z-1)d\cos(\theta^\mathrm{rb})+(l_y-1)d\sin(\theta^\mathrm{rb})\cos(\phi^\mathrm{rb})$.

It is important to note that the term $e^{j2\pi g\Delta f_{l}t},~l \in \mathcal{L}$ in received signal introduces a time-varying component into the beamforming process of the FD-RIS. This temporal fluctuation complicates both beamforming and signal transmission, thereby posing challenges for practical implementation.
To address this issue, a mixed-frequency technique combined with signal filtering is employed, as suggested in \cite{xu2018low, zhang2023distance}, to effectively suppress the time-dependent component $e^{j2\pi g\Delta f_{l}t}$. Specifically, the received signal is first amplified using a low-noise amplifier (LNA), and then mixed with the carrier frequency $f_\mathrm{c}$. The resulting intermediate-frequency signal is subsequently passed through a low-pass filter to extract the desired signal component. Hence, the final obtained signal can be expressed as
 \vspace{-1mm} \begin{align}\label{eq_received_signal}
  	y(t)=&\rho(d^\mathrm{ar})\rho(d^\mathrm{rb})s(t)\sum_{l_y=1}^{L_y}\sum_{l_z=1}^{L_z}e^{-j2\pi f_\mathrm{c}\frac{d^\mathrm{ar}_{l_y, l_z}}{c}}\times\notag\\
  	&A_0e^{j(\phi_0-2\pi g\Delta f_l \kappa_l)}e^{-j2\pi (f_\mathrm{c}+g\Delta f_l)}\frac{d^\mathrm{rb}_{l_y, l_z}}{c}+n_\mathrm{b}(t).
  \end{align}

Next, we will assess the beamforming capacity of the FD-RIS. Specifically, we temporarily ignore the large-scale path loss. According to the expression of the received signal in \eqref{eq_received_signal}, the normalized beamforming gain can be derived as \cite{wang2015frequency}
 \begin{align}
	BP=\frac{1}{L^2}\Big|\sum_{l_y=1}^{L_y}\sum_{l_z=1}^{L_z}e^{-j(\phi_{1, ly, l_z}-\phi_{2, ly, l_z}+\phi_{3, ly, l_z})}\Big|^2,
\end{align}
where $\phi_{1, ly, l_z}= 2\pi f_\mathrm{c}\frac{\Upsilon^\mathrm{ar}_{l_y, l_z}+\Upsilon^\mathrm{rb}_{l_y, l_z}}{c}$, $\phi_{2, ly, l_z}=\phi_0-2\pi g\Delta f_l \kappa_l$, $\phi_{3, ly, l_z}=2\pi g\Delta f_l\frac{d^\mathrm{rb}_{l_y, l_z}}{c}$. Note that to maximize the beamforming gain at Bob, an appropriate selection of the time delay, $\kappa_l$, can be made to ensure that $\phi_{2, ly, l_z}$ satisfies the phase alignment condition $\phi_{2, ly, l_z} = \phi_{1, ly, l_z} + \phi_{3, ly, l_z}$ for $l\in\mathcal{L}$.
\begin{figure}[ht]
	\centering
	\includegraphics[scale=0.34]{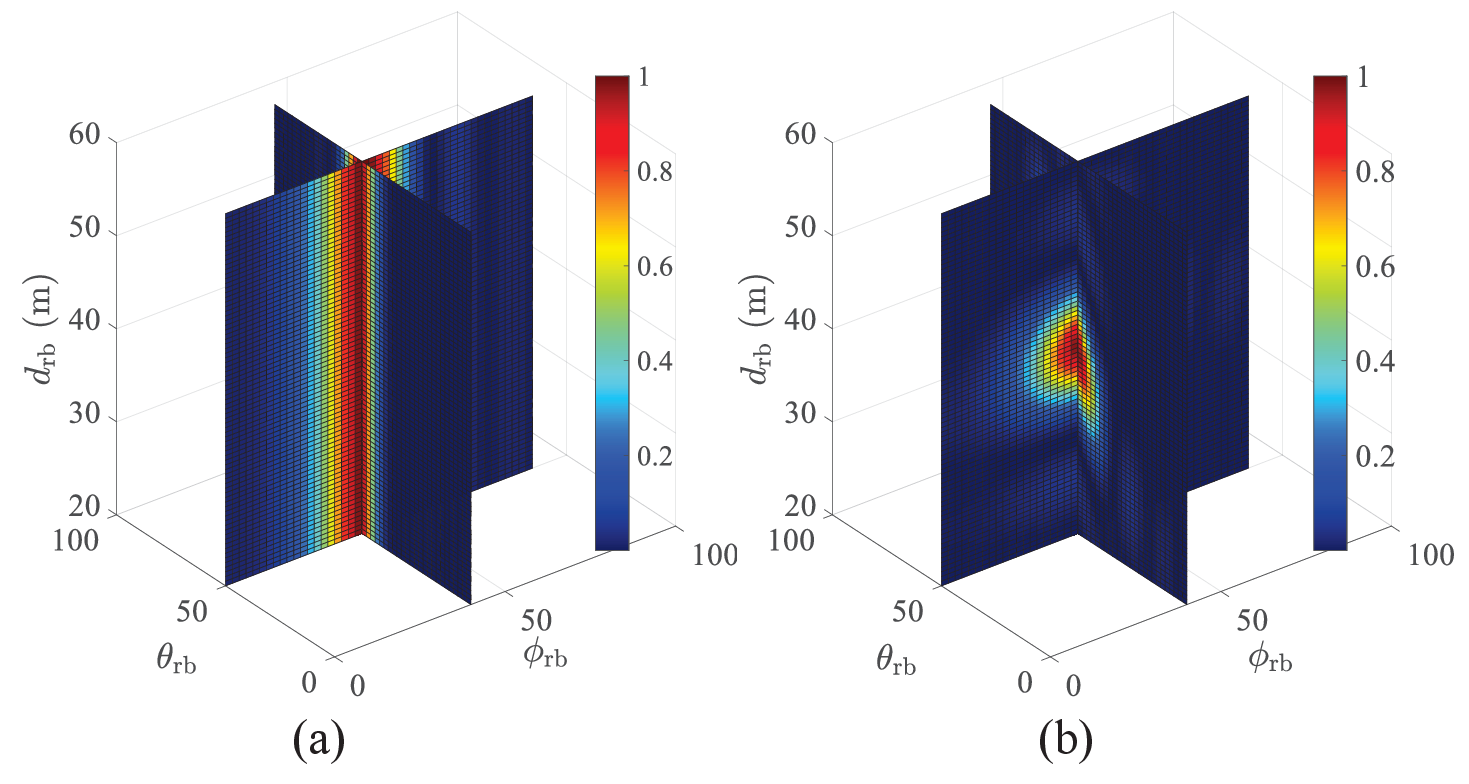}\\
	\caption{Beampattern considering $d_\mathrm{ar}=100$ m, $d_\mathrm{rb}=40$ m, $\theta_\mathrm{ar}=30^\circ$, $\phi_\mathrm{ar}=70^\circ$, $\theta_\mathrm{rb}=50^\circ$, $\phi_\mathrm{rb}=40^\circ$, $L=10\times 10$, $f_\mathrm{c}=28$ GHz, $\Delta f_l=10+ (l-1)\times30/(L-1)$ MHz: (a) Conventional RIS; (b) FD-RIS }\label{fig:beampattern_FD-RIS_conventional_RIS}
\end{figure}

Based on the beamforming gain model, Fig. \ref{fig:beampattern_FD-RIS_conventional_RIS} shows the beampattern comparison of  the conventional RIS and FD-RIS by taking account of maximizing Bob's beamforming gain.\footnote{The beamforming gain of the traditional RIS can be derived as: $BP_\mathrm{RIS}=\frac{1}{L^2}\left|\sum_{l_y=1}^{L_y}\sum_{l_z=1}^{L_z}e^{-j\phi_{1, ly, l_z}}U_{l_y, l_z}\right|^2$, where $U_{l_y, l_z}$ denotes the reflection coefficient of the $(l_y, l_z)$-th, which remains constant over time. To maximize the Bob's beamforming gain, $U_{l_y, l_z}$ should be configured to satisfy the condition: $U_{l_y, l_z}=e^{j\phi_{1, ly, l_z}}$.}
The results demonstrate that conventional RIS can only manipulate signal beamforming in the angular domain. However, this single-dimensional control strategy may lead to ``\textit{security blind zone}'' in communication systems. Specifically, \textcolor{blue}{ due to the one-dimensional beamforming capability of conventional RIS, the reflected signal energy can only be concentrated within the spatial region close to the legitimate user's direction. Considering the large-scale path loss, a high-energy region inevitably exists between RIS and the legitimate user. 
Once an illegitimate user enters this region, its' received signal power from the RIS may even exceed that of the legitimate user.
This phenomenon poses a significant challenge to ensuring the performance of CC, as it increases the risk of information leakage and undermines the confidentiality of the legitimate transmission. This high-power region is referred to as the ``\textit{security blind zone}'' of the entire communication system.
}

In contrast, FD-RIS possesses the capability to control signals in both the distance and angular domains though adjusting, offering new opportunities to break up the ``\textit{security blind zone}'' and enhance communication system concealment. By leveraging dual-domain control, FD-RIS can focus the signal energy precisely on the legitimate user’s position while significantly reducing the energy leakage in the region between the RIS and the user.
Hence, FD-RIS has the potential to overcome the challenge of security blind zone in RIS-aided systems. In the following sections, we will conduct a comprehensive analysis and numerical evaluations to demonstrate the effectiveness of FD-RIS in enhancing CC performance under various system settings especially in adversarial conditions.

\section{System Model and Analysis on CCs}\label{sec:S3}
\subsection{System Model}
\begin{figure}[ht]
	\centering
	\includegraphics[scale=0.36]{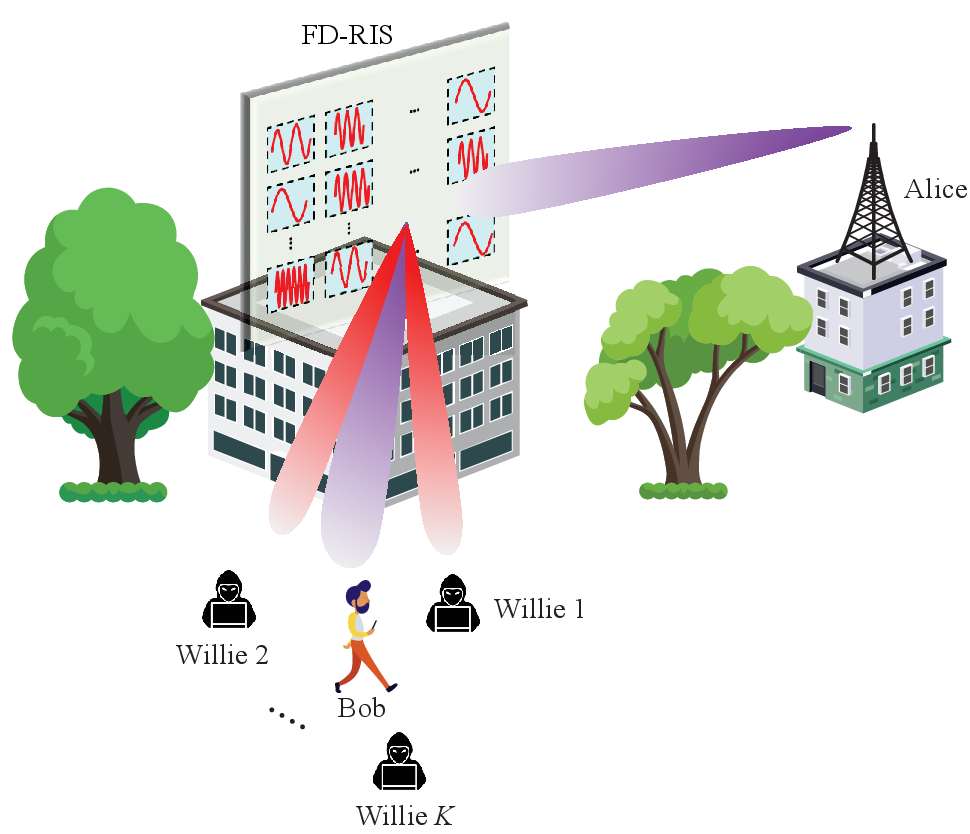}\\
	\caption{The system model of FD-RIS-assisted CCs.}\label{fig:scenario}
\end{figure}

Fig. \ref{fig:scenario} illustrates the system model of FD-RIS-assisted CCs, which consists of a single-antenna base station (Alice), a FD-RIS with $L=L_y\times L_z$ elements deployed near covert user (Bob), and $K$ potential wardens (Willie 1 to Willie $K$) distributed around Bob. These wardens aim to detect the communication behaviors between Alice and Bob in preparation for potential security attacks. It is assumed that all users, including Bob and the wardens, are equipped with a single antenna. Additionally, due to the presence of blockages such as buildings and trees, there are no direct links between Alice and any of the users, meaning that all transmissions from Alice to users should rely on the FD-RIS \footnote{\textcolor{blue}{We would like to emphasize that the proposed scheme can still be applied in scenarios with the line-of-sight (LoS) channel between transceivers, as the additional propagation paths can be incorporated into the overall channel model. In that case, the received signal would be the superposition of the RIS-reflected and the LoS components, and the beamforming optimization process can be naturally extended to jointly consider all paths.}}. 

 Considering the existence of line of sight (LoS) links between Alice and the FD-RIS, as well as the FD-RIS and Bob, we utilize the Rician channel model to characterize the connections between the FD-RIS and Alice/Bob, i.e. $\mathbf{h}_\mathrm{ar}$ and $\mathbf{h}_\mathrm{rb}$. Specifically, $\mathbf{h}_\mathrm{ar}$ and $\mathbf{h}_\mathrm{rb}$ can be expressed as
 \begin{align}
 \hspace{-2mm}	\mathbf{h}_{\eta}&=\rho(d^{\eta})\left(\beta_1\mathbf{h}_{\eta}^\mathrm{LoS}+\beta_2\mathbf{h}_{\eta}^\mathrm{NLoS}\right) \in\mathbb{C}^{L\times1},~\eta\in\{\mathrm{ar}, \mathrm{rb}\},
 \end{align}
where $\beta_1=\sqrt{\frac{\beta}{\beta+1}}$ and $\beta_2=\sqrt{\frac{1}{\beta+1}}$ with $\beta$ denoting the Rician factor. $\mathbf{h}_{\eta}^\mathrm{NLoS}$ is the NLoS component of $\mathbf{h}_{\eta}$, whose each element follows the  complex Gaussian distribution with zero mean and unit variance. According to the signal processing model, the LoS components can be given by
\begin{itemize}
	\item $\mathbf{h}_\mathrm{ar}^\mathrm{LoS}=\big[1, \cdots, e^{-j2\pi f_\mathrm{c}\Upsilon^\mathrm{ar}_{l_y, l_z}/c}, \cdots,e^{-j2\pi f_\mathrm{c}\Upsilon^\mathrm{ar}_{L_y, L_z}/c} \big]^T,$
	\item $
	\mathbf{h}_\mathrm{rb}^\mathrm{LoS}=\Big[e^{j2\pi\frac{g\Delta f_1 d_{1, 1}^\mathrm{rb}}{c} }, \cdots, e^{j2\pi\frac{f_\mathrm{c}\Upsilon^\mathrm{rb}_{l_y, l_z}+g\Delta f_l d_{l_y, l_z}^\mathrm{rb}}{c} }, \cdots,\notag\\ e^{j2\pi\frac{f_\mathrm{c}\Upsilon^\mathrm{rb}_{L_y, L_z}+g\Delta f_L d_{L_y, L_z}^\mathrm{rb}}{c} }\Big]^T.$
\end{itemize}

Building on the received signal model in \eqref{eq_received_signal} and the adopted channel model, the signal received by Bob is given by
\begin{align}
	y_\mathrm{b}[m]=\sqrt{P_\mathrm{t}}\mathbf{h}_\mathrm{rb}^H\boldsymbol{\Theta}\boldsymbol{\Theta}_0\mathbf{h}_\mathrm{ar}s[m]+n_\mathrm{b}[m],
\end{align}
where
\begin{itemize}
	\item $\boldsymbol{\Theta}=\operatorname{Diag}\{\vartheta_1, \cdots, \vartheta_l,\cdots, \vartheta_L\}$ denotes the passive beamforming matrix introduced by the time delays, where $\vartheta_l=e^{-j2\pi g \Delta f_l\kappa_l}, ~l\in\mathcal{L}$.
	\vspace{1mm}
	\item $\boldsymbol{\Theta}_0=A_0e^{j\phi_0}\mathbf{I}_{L\times L}$ is the initial reflection coefficient of the FD-RIS.
	\item $m\in\mathcal{M}=\{1, \cdots, M\}$ represents the index of channel uses with the maximum number of $M$ in a time slot.
	\item $s[m]\sim\mathcal{CN}(0, 1)$ denotes the transmitted signal from Alice to Bob.
\end{itemize}
\subsection{Analysis on CC}
In this paper, we will develop a CC scheme from the perspective of extreme scenarios with multiple wardens, to ensure that the proposed scheme can effectively withstand adverse conditions, 
making it more reliable and adaptable to challenging situations in real-world applications \cite{xiao2024simultaneously, xiao2024star-ris}. Specifically, it is assumed that the wardens (Willie $1$ - Willie $K$) in the considered scenario are highly capable and have access to the channel state information (CSI) between the FD-RIS and both Alice and Bob, as well as the system parameters, including the time-modulation frequency. Furthermore, each warden will employ the same signal processing techniques as Bob to detect any communications between Alice and Bob. Note that when Willie $k,~k\in\mathcal{K}\triangleq\{1, \cdots, K\}$, tries to detect the existence of communications between Alice and Bob, it faces a binary hypothesis testing problem based on the received signal sequences $\{y_{\mathrm{w}k}[m]\}_{m \in \mathcal{M}}$, i.e., null hypothesis $\mathcal{H}_{0, k}$ and alternative hypothesis $\mathcal{H}_{1, k}$. Specifically, the null hypothesis $\mathcal{H}_{0, k}$ represents the case where no communication exists between Alice and Bob, while the alternative hypothesis $\mathcal{H}_{1, k}$ indicates that Alice is transmitting to Bob. Under these two hypotheses, the received signal at Willie $k$ can be expressed as
\begin{align}
	y_{\mathrm{w}k}[m]=\begin{cases}
     n_{\mathrm{w}k}[m]	,&\mathcal{H}_{0, k},\\
	\sqrt{P_\mathrm{t}}\mathbf{h}_{\mathrm{rw}k}^H\boldsymbol{\Theta}\boldsymbol{\Theta}_0\mathbf{h}_\mathrm{ar}s[m]+n_{\mathrm{w}k}[m]	,& \mathcal{H}_{1, k},
	\end{cases}
\end{align}
where $\mathbf{h}_{\mathrm{rw}k}$ represents the channel between the FD-RIS and Willie $k$, which can be modeled similarly to the channel $\mathbf{h}_\mathrm{rb}$. $n_{\mathrm{w}k}[m]\sim\mathcal{CN}(0, \sigma^2_{\mathrm{w}k})$ is the AWGN at Willie $k$ with $\sigma^2_{\mathrm{w}k}$ denoting the noise power. Given that it is challenging to obtain the accurate $\sigma^2_{\mathrm{w}k}$ in practical communication scenarios, we introduce the noise power uncertainty at wardens\footnote{The noise uncertainty at Bob is neglected because it primarily affects the achievable rate, whereas the noise uncertainty at the wardens significantly influences wardens' detection capability}. The probability density function (PDF) of the noise power can be expressed as \cite{he2017covert}
\begin{align}
	f_{\sigma^2_{\mathrm{w}k}}(x)=\begin{cases}
		\frac{1}{2x\ln\varsigma}, &~\frac{\widehat{\sigma}^2_{\mathrm{w}k}}{\varsigma}\leq x\leq\varsigma\widehat{\sigma}^2_{\mathrm{w}k},\\
		0, &~otherwise,
	\end{cases}
\end{align}
where $\varsigma>1$ is a quantified parameter measuring the degree of noise uncertainty and $\widehat{\sigma}^2_{\mathrm{w}k}$ is the nominal noise power.

Following \cite{xiao2024star, Wang2025star-ris}, Willie $k$ employs the average received signal power across $M$ channel users, i.e., $\bar{P}_{\mathrm{w}k}=\frac{1}{M}\sum_{m=1}^{M}\left|y_{\mathrm{w}k}[m]\right|^2, ~ k\in\mathcal{K}, $ to conduct the statistical hypothesis test. 
By comparing $\bar{P}_{\mathrm{w}k}$ against a predefined detection threshold, i.e., $\tau_{\mathrm{dt},k}$, it establishes a decision rule to infer whether the transmission exists between Alice and Bob. The decision rule at Willie $k$ can be expressed as
\begin{align}
	\bar{P}_{\mathrm{w}k}=\frac{1}{M}\sum_{m=1}^{M}\left|y_{\mathrm{w}k}[m]\right|^2\underset{\mathcal{D}_{0, k}}{\stackrel{\mathcal{D}_{1, k}}{\gtrless}}\tau_{\mathrm{dt},k},~k\in\mathcal{K},
\end{align}
where $\mathcal{D}_{0, k}$ and $\mathcal{D}_{1, k}$ respectively denote the decisions that Willie $k$ tends to $\mathcal{H}_{0, k}$ and $\mathcal{H}_{1, k}$. Due to the fact that the value of the detection threshold $\tau_{\mathrm{dt},k}$ directly influences Willie $k$'s detection capabilities, we assume that Willie $k$ is able to properly design $\tau_{{\mathrm{dt}}, k}$ to maximize its detection accuracy. This process will undoubtedly make it more challenging for Alice and Bob to establish CCs.
Analogous to \cite{zheng2019multi}, we try to analyze Willie's detection ability by leveraging the large system analytic technique considering the asymptotic case of  $M\rightarrow\infty$. 
In this case, the average received power $\bar{P}_{\mathrm{w}k}$ can be asymptotically expressed as
\begin{align}
	\bar{P}_{\mathrm{w}k}=\begin{cases}
		\sigma^2_{\mathrm{w}k}, &~ \mathcal{H}_{0, k},\\
		P_\mathrm{t}\left|\mathbf{h}_{\mathrm{rw}k}^H\boldsymbol{\Theta}\boldsymbol{\Theta}_0\mathbf{h}_\mathrm{ar}\right|^2+\sigma^2_{\mathrm{w}k}, &~ \mathcal{H}_{1, k}.
	\end{cases}
\end{align}

In this paper, the detection error probability (DEP) is employed to assess Willies' ability for detecting the CCs between Alice and Bob. The DEP consists of two key components: false alarm (FA) probability and miss detection (MD) probability. The FA probability, denoted as $P_\mathrm{FA}$, refers to the likelihood that Willies incorrectly decide $\mathcal{D}_1$ under hypothesis $\mathcal{H}_0$, i.e., $P_\mathrm{FA}=\operatorname{Pr}(\mathcal{D}_1|\mathcal{H}_0)$. Conversely, the MD probability, denoted as $P_\mathrm{MD}$, represents the probability that Willies erroneously choose $\mathcal{D}_0$ under hypothesis $\mathcal{H}_1$, i.e., $P_\mathrm{MD}=\operatorname{Pr}(\mathcal{D}_0|\mathcal{H}_1)$. In particular, the DEP at Willie $k$ can be expressed as
\begin{align}
	P_{\mathrm{e}, k}&=P_{\mathrm{FA}, k}+P_{\mathrm{MD}, k}\notag\\
	&=\operatorname{Pr}(\mathcal{D}_{1, k}|\mathcal{H}_{0, k}) +\operatorname{Pr}(\mathcal{D}_{0, k}|\mathcal{H}_{1, k})\notag\\
	&=\operatorname{Pr}(\sigma^2_{\mathrm{w}, k}>\tau_{{\mathrm{dt}}, k}) +\notag\\
	&~~~\operatorname{Pr}(P_\mathrm{t}\left|\mathbf{h}_{\mathrm{rw}k}^H\boldsymbol{\Theta}\boldsymbol{\Theta}_0\mathbf{h}_\mathrm{ar}\right|^2+\sigma^2_{\mathrm{w}k}<\tau_{{\mathrm{dt}}, k}).
\end{align}
According to the PDF of $\sigma^2_{\mathrm{w}k}$, the DEP of Willie $k$ can be derived as
\begin{align}\label{eq_DEP}
	P_{\mathrm{e}, k}=&1-\operatorname{Pr}(\tau_{{\mathrm{dt}}, k}-\omega_k\leq\sigma^2_{\mathrm{w}k}\leq\tau_{{\mathrm{dt}}, k})\notag\\
	\stackrel{(a)}{=}&1-\int_{\max\Big\{\tau_{{\mathrm{dt}}, k}-\omega_k, \frac{\widehat{\sigma}^2_{\mathrm{w}k}}{\varsigma}\Big\}}^{\tau_{{\mathrm{dt}}, k}}\frac{1}{2x\ln\varsigma} dx\notag\\
	=&1-\frac{\left(\ln\tau_{{\mathrm{dt}}, k}-\ln\Big(\max\Big\{\tau_{{\mathrm{dt}}, k}-\omega_k, \frac{\widehat{\sigma}^2_{\mathrm{w}k}}{\varsigma}\Big\}\Big)\right)}{2\ln\varsigma},
\end{align}
where $\omega_k=P_\mathrm{t}\left|\mathbf{h}_{\mathrm{rw}k}^H\boldsymbol{\Theta}\boldsymbol{\Theta}_0\mathbf{h}_\mathrm{ar}\right|^2$.
Here, $(a) $ follows from the assumption that $\tau_{{\mathrm{dt}}, k}  \in [\frac{\widehat{\sigma}^2_{\mathrm{w}k}}{\varsigma},~ \varsigma\widehat{\sigma}^2_{\mathrm{w}k}]$, where the range is carefully chosen to enhance Willies' detection capability, thereby ensuring that the CC scheme developed in the following sections remains both robust and practically viable under various challenging conditions. Notably, when $\tau_{{\mathrm{dt}}, k} < \frac{\widehat{\sigma}^2_{\mathrm{w}k}}{\varsigma} ~ \text{or} ~ \tau_{{\mathrm{dt}}, k} > \omega_k + \varsigma\widehat{\sigma}^2_{\mathrm{w}k},$ the DEP satisfies $P_{\mathrm{e}, k} = 1$ at all times. This indicates that Willie \( k \) is completely incapable of accurately detecting the communications between Alice and Bob. Furthermore, when $\tau_{{\mathrm{dt}}, k} >\varsigma\widehat{\sigma}^2_{\mathrm{w}k}$, the FA probability remains at $P_{\mathrm{FA}, k} = 0$, while the MD probability $P_{\mathrm{MD}, k}$ increases as $ \tau_{{\mathrm{dt}}, k} $ increases, and finally asymptotically approaches 1. This suggests that in such cases, Willie $k$ is increasingly likely to overlook ongoing communication.

Recall that wardens will select suitable detection thresholds $\{\tau_{{\mathrm{dt}}, k} \}_{k\in\mathcal{K}}$ to maximize their detection abilities or minimize their DEP for CCs between Alice and Bob. Next, we will derive the optimal detection thresholds  and the  optimal DEP based on the DEP expression in \eqref{eq_DEP}. Specifically, the closed-form optimal detection threshold of Willie $k$ can be obtained by Theorem \ref{th1}.
\begin{theorem}\label{th1}
The close-form optimal detection threshold $\tau_{{\mathrm{dt}}, k}^\mathrm{opt}$ for Willie $k$ can be derived as
\begin{align}\label{eq_opt_threshold}
	\tau_{{\mathrm{dt}}, k}^\mathrm{opt}=\min\left\{\omega_k+\frac{\widehat{\sigma}^2_{\mathrm{w}k}}{\varsigma},~ \varsigma\widehat{\sigma}^2_{\mathrm{w}k}\right\}.
\end{align}
	\begin{proof}
		Proof is presented in Appendix \ref{appedix 1}.
	\end{proof}
\end{theorem}
Substituting \eqref{eq_opt_threshold} into \eqref{eq_DEP}, the optimal DEP for Willie $k$ can be expressed as
\begin{align}\label{eq_opt_DEP}
	P^\mathrm{opt}_{\mathrm{e}, k}=\begin{cases}
	1-\frac{\ln\left(1+\frac{\varsigma\omega_k}{\widehat{\sigma}^2_{\mathrm{w} k}}\right)}{2\ln\varsigma}, &~ \omega_k\leq\frac{(\varsigma^2-1)\widehat{\sigma}^2_{\mathrm{w}k}}{\varsigma},\\
		0, &~ otherwise.
	\end{cases}
\end{align}

After deriving the optimal DEP from the perspective of the worst scenario, to guarantee the covertness of communications between Alice and Bob, $\omega_k\leq\frac{(\varsigma^2-1)\widehat{\sigma}^2_{\mathrm{w}k}}{\varsigma}$ and $P^\mathrm{opt}_{\mathrm{e}, k}\geq1-\xi$ must be satisfied simultaneously, where $\xi\in(0, 1)$ denotes the covert requirement of the communication systems, which is a sufficiently small value. Hence, we have
\begin{align}\label{eq_covert_constrain}
	\omega_k\leq\min\left\{\frac{(\varsigma^2-1)\widehat{\sigma}^2_{\mathrm{w}k}}{\varsigma},\frac{(\varsigma^{2\xi}-1)\widehat{\sigma}^2_{\mathrm{w}k}}{\varsigma}\right\} \stackrel{(b)}{=}\frac{(\varsigma^{2\xi}-1)\widehat{\sigma}^2_{\mathrm{w}k}}{\varsigma},
\end{align}
where $(b)$ is achieved due to the fact that $\xi\in(0, 1)$. 

It is important to highlight that the optimal DEP in \eqref{eq_opt_DEP} is derived from Willie $k$'s perspective. However, from Alice's standpoint, the parameter $\omega_k$ is inherently a random variable due to the presence of the NLoS component in $\mathbf{h}_{\mathrm{rw}k}$\footnote{The communication system under consideration is presumed to incorporate sensing capabilities \cite{su2023sensing}. As a result, Alice can leverage this functionality to determine the locations of potential wardens. The accuracy of wardens' localization is inherently dependent on Alice's sensing proficiency. In this context, \textcolor{blue}{we assume that Alice possesses strong sensing functionality, making it feasible to acquire accurate location information of potential wardens}. Accordingly, the LoS component of $\mathbf{h}_{\mathrm{rw}k}$ is assumed to be perfectly known to Alice. \textcolor{blue}{It is worth noting that when Alice has limited sensing capability regarding potential wardens, meaning that only a certain position range can be estimated, the proposed scheme can still be applied to this scenario with some simple extensions. More details can refer to \cite{zhang2023distance}.}
}. To address this challenge, existing approaches in the literature (e.g., \cite{xiao2024simultaneously, wang2021intelligent}) commonly approximate the covert constraint \eqref{eq_covert_constrain} by computing the average of $\omega_k$ over $\mathbf{h}_{\mathrm{rw}k}$, i.e., $\mathbb{E}_{\mathbf{h}_{\mathrm{rw}k}}(\omega_k)$, and then substituting this average value into the constraint formulation. However, merely ensuring that the average satisfies the condition, i.e., $\mathbb{E}_{\mathbf{h}{\mathrm{rw}k}}(\omega_k)\leq \frac{(\varsigma^{2\xi}-1)\widehat{\sigma}^2_{\mathrm{w}k}}{\varsigma}$, cannot guarantee that all possible realizations of $\omega_k$ will satisfy the constraint. In other words, designing the system parameters based on the average policy may neglect certain extreme cases, potentially leading to violations of the CC constraint in some scenarios. To address this challenge, the logarithmic moment generating function \cite{xiao2024adaptive} of $\omega_k$, denoted as $\tilde{\omega}_k$, is employed to approximate $\omega_k$ and subsequently incorporated into the covert constraint, where $\tilde{\omega}_k$ is defined as
\begin{align}
	\tilde{\omega}_k=\frac{1}{\psi}\log\mathbb{E}(e^{\psi\omega_k}),
\end{align}
where $\psi\geq 0$ represents the penalty exponent.
\begin{remark}
	\textcolor{blue}{In this section, we will analyze the effectiveness of employing $\tilde{\omega}_k$ to approximate $\omega_k$ based on the following two cases.}
		
\textcolor{blue}{	Case 1: when $\psi$ approaches $0$, we have
	\begin{align}
		\lim\limits_{\psi\rightarrow 0}\tilde{\omega}_k=&\lim\limits_{\psi\rightarrow 0}\frac{1}{\psi}\log\mathbb{E}(e^{\psi\omega_k})\notag\\
		\stackrel{(b)}{=}&\lim\limits_{\psi\rightarrow 0}\frac{\mathbb{E}(\omega_ke^{\psi\omega_k})}{\mathbb{E}(e^{\psi\omega_k})}=\mathbb{E}(\omega_k),
	\end{align}
where $(b)$ is due to the L'Hospital's rule. }

\textcolor{blue}{	Case 2: when $\psi$ approaches $+\infty$, we have
	\begin{align}
		\lim\limits_{\psi\rightarrow +\infty}\tilde{\omega}_k=\lim\limits_{\psi\rightarrow +\infty}\frac{1}{\psi}\log\mathbb{E}(e^{\psi\omega_k}).
	\end{align}
	Actually, due to the amplification characteristic of the exponential function, we can derive $\mathbb{E}(e^{\psi\omega_k})\approx e^{\psi\max(\omega_k)}$. Thus,
	\begin{align}
		\lim\limits_{\psi\rightarrow +\infty}\frac{1}{\psi}\log\mathbb{E}(e^{\psi\omega_k})\approx\frac{\psi\max(\omega_k)}{\psi}=\max(\omega_k).
	\end{align}}
	
 \textcolor{blue}{According to the derived results in these two cases, we can find that an appropriate selecting for $\psi$ can enable $\tilde{\omega}_k$ to effectively approximate the maximum value of $\omega_k$. Such an approximation facilitates a conservative yet reliable system design, which in turn significantly reduces the probability that the wardens successfully detect the legitimate communications between Alice and Bob.}
\end{remark}
 For the analytical expression of $\tilde{\omega}_k$ is given by the following theorem.
\begin{theorem}\label{th2}
According to distribution of $\omega_k$, the analytical expression of $\tilde{\omega}_k$ can be derived as
	\begin{align}
		\tilde{\omega}_k=&\frac{1}{\psi}\log\mathbb{E}(e^{\psi\omega_k})\notag\\
		=&\frac{|\mu_k|^2}{1-\psi\tilde{\sigma}^2_k}-\frac{\log(1-\psi\tilde{\sigma}^2_k)}{\psi}, ~~~  ~~\tilde{\sigma}^2_k\psi<1
	\end{align}
	where
	\begin{itemize}
		\item $\mu_k=\sqrt{P_\mathrm{t}}\rho(d^{\mathrm{rw}k})\beta_1(\mathbf{h}^\mathrm{LoS}_{\mathrm{rw}k})^H\boldsymbol{\Theta}\boldsymbol{\Theta}_0\mathbf{h}_\mathrm{ar}$,
		\vspace{1mm}
		\item $\tilde{\sigma}^2_k=P_\mathrm{t}\rho^2(d^{\mathrm{rw}k})\left\|\beta_2\boldsymbol{\Theta}\boldsymbol{\Theta}_0\mathbf{h}_\mathrm{ar}\right\|^2_2$.
	\end{itemize}

	\begin{proof}
		Proof is presented in Appendix \ref{appedix 2}.
	\end{proof}
\end{theorem}
	\textcolor{blue}{Note that due to the inclusion of large-scale path loss in $\tilde{\sigma}_k^2$, the magnitude of $\tilde{\sigma}_k^2$ is typically very small, and its order of magnitude is only minimally affected by variations in the optimal solution. This implies that there is a wide range of $\psi$ values that satisfy the condition $\tilde{\sigma}_k^2 \psi < 1$. Therefore,  to ensure that the condition is always met, we can select a reasonably large $\psi$ within this range. It is important to note that a larger $\psi$ is preferred because it helps $\tilde{\omega}_k$ better approximate the upper bound of $\omega_k$, thereby improving the approximation accuracy and analytical reliability.
	}
	
Therefore, based on \eqref{eq_covert_constrain} and the above analysis, we can utilize the covert constrain $\tilde{\omega}_k\leq\frac{(\varsigma^{2\xi}-1)\widehat{\sigma}^2_{\mathrm{w}k}}{\varsigma}$ to guarantee the CCs between Alice and Bob, which can be further expressed as
\begin{align}\label{eq_covert_constrain_trans}
|\mu_k|^2\leq\underbrace{(1-\psi\tilde{\sigma}^2_k)\left(\frac{(\varsigma^{2\xi}-1)\widehat{\sigma}^2_{\mathrm{w}k}}{\varsigma}+\log(1-\psi\tilde{\sigma}^2_k)\right)}_{h_k},
\end{align}
which can be interpreted as: To ensure CCs between Alice and Bob, the signal power transmitted through the LoS components of the eavesdropping channels to the wardens must be suppressed below a threshold determined by the CC requirements.
\begin{remark}
	Based on the covert constraint given in \eqref{eq_covert_constrain_trans}, several important observations can be made. Specifically, let us define $x_k = 1 - \psi \tilde{\sigma}_k^2 > 0$, from which the minimum value of the expression $h_k$ shown in  \eqref{eq_covert_constrain_trans} can be derived as $h_k^{\mathrm{opt}} = -e^{-\widehat{\mu}_k - 1}$, and the corresponding $x_k^{\mathrm{opt}} = e^{-\widehat{\mu}_k - 1} > 0$, where $\widehat{\mu}_k = \frac{(\varsigma^{2\xi} - 1)\widehat{\sigma}{\mathrm{w}k}^2}{\varsigma}$.
	It is worth noting that $h_k^{\mathrm{opt}} < 0$ always holds under this formulation. This implies that, in certain extreme cases, the term $|\mu_k|^2$ would need to be negative to satisfy the constraint. However, this leads to a contradiction, as $|\mu_k|^2 \geq 0$ must always be satisfied from a physical and mathematical standpoint.
	Therefore, the covert constraint in \eqref{eq_covert_constrain_trans} may become infeasible under specific parameter settings. To mitigate this, we reformulate the covert constraint as
	\begin{align}\label{eq_covert_constrain_trans_}
		|\mu_k|^2\leq \max(0, h_k),~ k\in\mathcal{K}.
	\end{align}
	Specifically, when $h_k<0$, we enforce $|\mu_k|^2=0$ through the design of time delays and modulation frequencies.
\end{remark}

Furthermore, based on the received signal model, the achievable communication rate at Bob is given as
\begin{align}\label{eq_R_b}
	R_\mathrm{b}=\log\left(1+P_\mathrm{t}\left|\mathbf{h}_\mathrm{rb}^H\boldsymbol{\Theta}\boldsymbol{\Theta}_0\mathbf{h}_\mathrm{ar}\right|^2/\sigma^2_\mathrm{b}\right).
\end{align}
\section{Problem Formulation and Algorithm Design}\label{sec:S4}
\subsection{Problem Formulation}
 To evaluate the effectiveness of the FD-RIS, an optimization problem is formulated to maximize Bob's covert rate, subject to covertness constraints, through joint optimization of modulation frequency and time-delay variables. Specifically, the formulated optimization problem is expressed as follows:
\begin{subequations}\label{eq_ori_opt}
	\begin{align}
		&\max _{\boldsymbol{\kappa},~ \mathbf{f}} ~ R_\mathrm{b},\notag \\
		&~\text { s.t. } \Delta f_{\min}\leq\Delta f_l\leq \Delta f_{\max},~ l\in\mathcal{L},\label{eq_ori_opt_1}\\
		&\qquad 0\leq\kappa_l\leq\	\frac{1}{\Delta f_l}, ~l\in\mathcal{L}, \label{eq_ori_opt_2}\\
		&\qquad |\mu_k|^2\leq \max(0, h_k), ~k\in\mathcal{K}, \label{eq_ori_opt_3}
	\end{align}
\end{subequations}
where $\boldsymbol{\kappa}\triangleq\{\kappa_l\}_{l\in\mathcal{L}}$, $\mathbf{f}\triangleq\{\Delta f_l\}_{l\in\mathcal{L}}$ denote the sets of time delay and modulation frequency. $\Delta f_{\max}$ and $\Delta f_{\min}$ are the upper and lower bounds of modulation frequency. It is worth noting that addressing this problem is significantly difficult due to the non-convexity of objective function and covert constrains, as well as the strong coupling among variables. To address this challenge, we employ an alternative strategy that decomposes the optimization problem into two subproblems, facilitating independent design of time delays and modulation frequencies, respectively. Next, we will design the corresponding algorithms to effectively solve these two subproblems.

\subsection{Algorithm Design}
\subsubsection{Passive Beamforming Design for Time Delays}This section focuses on optimizing the time-delay variables, given the modulation frequency variables in $\mathbf{f}$. Hence, the original optimization problem in \eqref{eq_ori_opt} can be simplified as follows:
\begin{subequations}\label{eq_time_delay}
	\begin{align}
		&\max _{\boldsymbol{\kappa}} ~R_\mathrm{b},\notag \\
		&~\text { s.t. } 0\leq\kappa_l\leq\	\frac{1}{\Delta f_l}, ~l\in\mathcal{L}, \label{eq_time_delay_1}\\
		&\qquad |\mu_k|^2\leq \max(0, h_k), ~k\in\mathcal{K}.\label{eq_time_delay_2}
	\end{align}
\end{subequations}

\begin{remark}
It is noteworthy that directly optimizing the time-delay variables $\kappa_l,~l \in \mathcal{L}$ is challenging, as they are intricately coupled in the expression $\vartheta_l = e^{-j2\pi g \Delta f_l\kappa_l}$. To overcome this difficulty, we adopt a holistic design strategy by optimizing the reflection coefficients $\vartheta_l$ induced by the time delays, rather than the time delays themselves. This approach is justified by the fact that, for any given $\Delta f_l$, the corresponding time delays can always be retrieved from the optimized values of $\vartheta_l$ via the inverse mapping of the exponential relation.
\end{remark}

Thus, the optimization problem \eqref{eq_time_delay} can be equivalently transformed as
\vspace{-2mm}
\begin{subequations}\label{eq_time_delay_trans}
	\begin{align}
		&\max _{
		\boldsymbol{\vartheta}} ~R_\mathrm{b},\notag \\
		&~\text { s.t. }\left|\mathcal{I}_l(\boldsymbol{\vartheta})\right|=1, ~l\in\mathcal{L}, \label{eq_time_delay_trans_1}\\
		&\qquad |\mu_k|^2\leq \max\Big(0, h_k\Big), ~k\in\mathcal{K},\label{eq_time_delay_trans_2}
	\end{align}
\end{subequations}
where $\boldsymbol{\vartheta}=\operatorname{diag}(\boldsymbol{\Theta}^{*})$. Actually, solving problem \eqref{eq_time_delay_trans} is still challenging because of the non-concave objective function and constant modulus constraints in \eqref{eq_time_delay_trans_1}. 
 To tackle this issue, we first utilize the MMSE \cite{xiaorubust2025} method to equivalently transformed $R_b$ as
\begin{align}\label{Rb_fb}
	R_b=\frac{1}{\ln2}\max_{W_\mathrm{b}, u_\mathrm{b}}\underbrace{\ln(W_\mathrm{b})-W_\mathrm{b}E_\mathrm{b}(\boldsymbol{\vartheta}, u_\mathrm{b})+1}_{f_\mathrm{b}},
\end{align}
where $E_\mathrm{b}=\left(\sqrt{P_\mathrm{t}}u_\mathrm{b}^H\boldsymbol{\vartheta}^H\tilde{\mathbf{h}}-1\right)\left(\sqrt{P_\mathrm{t}}u_\mathrm{b}^H\boldsymbol{\vartheta}^H\tilde{\mathbf{h}}-1\right)^H+\sigma_\mathrm{b}^2u_\mathrm{b}^Hu_\mathrm{b}$ with $\tilde{\mathbf{h}}=\mathbf{h}_\mathrm{rb}^*\circ(\boldsymbol{\Theta}_0\mathbf{h}_\mathrm{ar})$. Note that the expression $f_\mathrm{b}$ shown in \eqref{Rb_fb} is a concave function w.r.t. variables $u_\mathrm{b}$ and $W_\mathrm{b}$. Thus, the optimal $u_\mathrm{b}$ and $W_\mathrm{b}$ can be determined by nullifying the first-order partial derivative of $f_\mathrm{b}$ w.r.t. the relevant parameters, which yields the following solutions:
\begin{align}\label{eq_mmse_aux_varibale}
	u_\mathrm{b}^\mathrm{opt}=\frac{\sqrt{P_\mathrm{t}}\boldsymbol{\vartheta}^H\tilde{\mathbf{h}}}{\left|\sqrt{P_\mathrm{t}}\boldsymbol{\vartheta}^H\tilde{\mathbf{h}}\right|^2+\sigma_\mathrm{b}^2}, ~W^\mathrm{opt}_\mathrm{b}=E^{-1}.
\end{align}

\textcolor{blue}{It is worth noting that the equivalent relationship between the logarithmic form in \eqref{eq_R_b} and the right side of \eqref{Rb_fb}  can be demonstrated by substituting $u_\mathrm{b}^\mathrm{opt}$ and $W_\mathrm{b}^\mathrm{opt}$ into the objective function of \eqref{Rb_fb}.}
on the basis of the discussion above, we can obtain the concave lower-bound of $R_b$ in $(p+1)$-th iteration of the iterative algorithm, which is given by
\begin{align}
		R_b\geq\tilde{R}^{(p)}_\mathrm{b}=&\frac{1}{\ln2}\left(\ln(W^{(p)}_\mathrm{b})-W_\mathrm{b}E_\mathrm{b}(\boldsymbol{\vartheta}, u_\mathrm{b}^{(p)})+1\right)\notag\\
		=&-\boldsymbol{\vartheta}^H\mathbf{A}^{(p)}\boldsymbol{\vartheta}+2\operatorname{Re}(\boldsymbol{\vartheta}^H\mathbf{a}^{(p)})+\tilde{c}^{(p)},
\end{align}
where
\begin{itemize}
	\item  $\mathbf{A}^{(p)}=\frac{P_\mathrm{t}W^{(p)}_\mathrm{b}\tilde{\mathbf{h}}(u^{(p)}_\mathrm{b})^Hu^{(p)}_\mathrm{b}\tilde{\mathbf{h}}^H}{\ln2}, ~\mathbf{a}^{(p)}=\frac{\sqrt{P_\mathrm{t}}W_\mathrm{b}^{(p)}\tilde{\mathbf{h}}(u^{(p)}_\mathrm{b})^H}{\ln2}$,
	\vspace{1mm}
	\item $\tilde{c}^{(p)}=\frac{-W_\mathrm{b}^{(p)}-W_\mathrm{b}^{(p)}\left|u_\mathrm{b}^{(p)}\right|^2+\ln W^{(p)}_\mathrm{b}}{\ln2}$.
\end{itemize}
It is worth noting that $W^{(p)}_\mathrm{b}$ and $u^{(p)}_\mathrm{b}$ are obtained by using the expressions in \eqref{eq_mmse_aux_varibale} and the obtained solution of $\boldsymbol{\vartheta}$ in the $p$-th iteration. Hence, in the $(p+1)$-th iteration, we can transform the optimization subproblem \eqref{eq_time_delay_trans} as
\begin{subequations}\label{eq_time_delay_trans_mmse}
	\begin{align}
		&\max _{\boldsymbol{\vartheta}} ~-\boldsymbol{\vartheta}^H\mathbf{A}^{(p)}\boldsymbol{\vartheta}+2\operatorname{Re}(\boldsymbol{\vartheta}^H\mathbf{a}^{(p)}),\notag \\
		&~\text { s.t. } \left|\mathcal{I}_l(\boldsymbol{\vartheta})\right|=1, ~l\in\mathcal{L}, \label{eq_time_delay_trans_mmse_1}\\
		&~\qquad \boldsymbol{\vartheta}^H\mathbf{B}_k\boldsymbol{\vartheta}-\iota_k\leq 0, ~k\in\mathcal{K},\label{eq_time_delay_trans_mmse_2}
	\end{align}
\end{subequations}
where the covert constraint \eqref{eq_time_delay_trans_2} is equivalently transformed into \eqref{eq_time_delay_trans_mmse_2} by defining
$\mathbf{B}_k=\left((\mathbf{h}^\mathrm{LoS}_{\mathrm{rw}k})^*\circ(\boldsymbol{\Theta}_0\mathbf{h}_\mathrm{ar})\right)((\mathbf{h}^\mathrm{LoS}_{\mathrm{rw}k})^*\circ$ $(\boldsymbol{\Theta}_0\mathbf{h}_\mathrm{ar}))^H$,  $\iota_k=\frac{\max(0, h_k)}{P_\mathrm{t}\rho^2(d^{\mathrm{rw}k})\beta^2_1}$. To tackle the constant constraints in \eqref{eq_time_delay_trans_mmse_1}, the penalty dual decomposition (PDD) algorithm framework \cite{shi2020penalty} is leveraged to handle this non-convex optimization problem. Specifically, an auxiliary variable $\widehat{\boldsymbol{\vartheta}}$
is first introduced to equivalently convert \eqref{eq_time_delay_trans_mmse} as
\begin{subequations}\label{eq_time_delay_trans_PDD}
	\begin{align}
		&\max _{\boldsymbol{\vartheta},~ \widehat{\boldsymbol{\vartheta}}} ~-\boldsymbol{\vartheta}^H\mathbf{A}^{(p)}\boldsymbol{\vartheta}+2\operatorname{Re}(\boldsymbol{\vartheta}^H\mathbf{a}^{(p)}),\notag \\
		&~\text { s.t. } \left\|\boldsymbol{\vartheta}\right\|_2^2\leq L,\label{eq_time_delay_trans_PDD_1}\\
		&~\qquad \boldsymbol{\vartheta}=\widehat{\boldsymbol{\vartheta}},\label{eq_time_delay_trans_PDD_2}\\
		&~\qquad  \left|\mathcal{I}_l(\widehat{\boldsymbol{\vartheta}})\right|=1, ~l\in\mathcal{L},\label{eq_time_delay_trans_PDD_3} \\
		&~\qquad \boldsymbol{\vartheta}^H\mathbf{B}_k\boldsymbol{\vartheta}-\iota_k\leq 0, ~k\in\mathcal{K}.\label{eq_time_delay_trans_PDD_4}
	\end{align}
\end{subequations}
Subsequently, the augmented Lagrangian method is employed to reformulate the equality constraint \eqref{eq_time_delay_trans_PDD_2}, incorporating it into the objective function as a penalty term. Therefore the augmented Lagrangian problem of \eqref{eq_time_delay_trans_PDD} can be expressed as
\begin{subequations}\label{eq_time_delay_trans_PDD_al}
	\begin{align}
		&\max _{\boldsymbol{\vartheta},~ \widehat{\boldsymbol{\vartheta}}} ~-\boldsymbol{\vartheta}^H\mathbf{A}^{(p)}\boldsymbol{\vartheta}+2\operatorname{Re}(\boldsymbol{\vartheta}^H\mathbf{a}^{(p)})-\frac{1}{2\breve{\rho}}\left\|\boldsymbol{\vartheta}-\widehat{\boldsymbol{\vartheta}}+\breve{\rho}\boldsymbol{\lambda}\right\|^2_2,\notag \\
		&~\text { s.t. }~ \eqref{eq_time_delay_trans_PDD_1},~ \eqref{eq_time_delay_trans_PDD_3},~ \eqref{eq_time_delay_trans_PDD_4},\label{eq_time_delay_trans_PDD_al_1}
	\end{align}
\end{subequations}
where $\breve{\rho}$ denotes the penalty coefficient associated with the equality constrain in \eqref{eq_time_delay_trans_PDD_2}, $\boldsymbol{\lambda}$ is the set of the Lagrange multipliers. It is important to highlight that the PDD method constitutes a nested iterative procedure. The inner loop employs an alternating strategy to solve the augmented Lagrangian problem, while the outer loop updates either the Lagrange multipliers or the penalty coefficients to guarantee the equality constraints. In particular, we first utilize the alternative strategy to solve the augmented Lagrangian problem \eqref{eq_time_delay_trans_PDD_al}. Specifically, when variable $\widehat{\boldsymbol{\vartheta}}$ is given, the augmented Lagrangian problem can be simplified as
\begin{subequations}\label{eq_time_delay_trans_PDD_al1}
	\begin{align}
		&\max _{\boldsymbol{\vartheta}} ~-\boldsymbol{\vartheta}^H\mathbf{A}^{(p)}\boldsymbol{\vartheta}+2\operatorname{Re}(\boldsymbol{\vartheta}^H\mathbf{a}^{(p)})-\frac{1}{2\breve{\rho}}\left\|\boldsymbol{\vartheta}-\widehat{\boldsymbol{\vartheta}}+\breve{\rho}\boldsymbol{\lambda}\right\|^2_2,\notag \\
		&~\text { s.t. }~ \eqref{eq_time_delay_trans_PDD_1},~ \eqref{eq_time_delay_trans_PDD_4}.\label{eq_time_delay_trans_PDD_al1_1}
	\end{align}
\end{subequations}
Note that the optimization problem \eqref{eq_time_delay_trans_PDD_al1} is convex, therefore, the existing optimization solvers such as CVX can be employed to effectively  address it.

After obtaining variable $\boldsymbol{\vartheta}$, we focus on designing $\widehat{\boldsymbol{\vartheta}}$. The subproblem associated with variable $\widehat{\boldsymbol{\vartheta}}$ can be expressed as
\begin{subequations}\label{eq_time_delay_trans_PDD_al2}
	\begin{align}
		&\max _{\widehat{\boldsymbol{\vartheta}}} ~-\frac{1}{2\breve{\rho}}\left\|\boldsymbol{\vartheta}-\widehat{\boldsymbol{\vartheta}}+\breve{\rho}\boldsymbol{\lambda}\right\|^2_2,\notag \\
		&~\text { s.t. }  \left|\mathcal{I}_l(\widehat{\boldsymbol{\vartheta}})\right|=1, ~l\in\mathcal{L}.\label{eq_time_delay_trans_PDD_al2_1}
	\end{align}
\end{subequations}
In terms of this problem, we can derived the optimal $\widehat{\boldsymbol{\vartheta}}$	as
$\label{eq_opt_auxi}
	\widehat{\boldsymbol{\vartheta}}=e^{j\angle(\boldsymbol{\vartheta}+\breve{\rho}\boldsymbol{\lambda})}
$ by adopting the phase shift alignment method.
\begin{center}
	\begin{tabular}{p{8.5cm}}
		\toprule[2pt]
		\textbf{Algorithm 1:}  PDD method for Solving Subproblem \eqref{eq_time_delay_trans_mmse}   \\
		\midrule[1pt]
		1: Initialize the Lagrange multiplier $\boldsymbol{\lambda}^{(0)}$, penalty coeffi-\\
		\quad cient $\breve{\rho}$ and tolerance accuracy $\tilde{\epsilon}$; Set iteration index $v=$0.\\
		2: \textbf{Repeat} \\
		3:\quad \textbf{Repeat} \\
		4: \qquad Solving subproblem \eqref{eq_time_delay_trans_PDD_al1} with given $\widehat{\boldsymbol{\vartheta}}$.\\
		5: \qquad Updating $\widehat{\boldsymbol{\vartheta}}$ with obtained $\boldsymbol{\vartheta}$.\\
		6:\quad \textbf{Until} the difference in objective function values betw-\\
		\qquad een adjacent iterations falls below predefined threshold.\\
		7:\quad Obtain $\left(\boldsymbol{\vartheta}^{(v+1)}, \widehat{\boldsymbol{\vartheta}}^{(v+1)}\right)$ and calculate $\breve{d}^{(v+1)}=$\\
		\qquad $\|\boldsymbol{\vartheta}^{(v+1)}-\widehat{\boldsymbol{\vartheta}}^{(v+1)}\|_2$.\\
		8: \quad if $\breve{d}^{(v+1)}\leq\epsilon^{(v+1)}$, $\boldsymbol{\lambda}^{(v+1)}\leftarrow \boldsymbol{\lambda}^{(v)}+\frac{\boldsymbol{\vartheta}^{(v+1)}-\widehat{\boldsymbol{\vartheta}}^{(v+1)}}{\breve{\rho}}$.\\
		9: \quad Otherwise, $\breve{\rho}\leftarrow \breve{\xi}\breve{\rho}$. Let $v\leftarrow v+1$.\\
		10:~\textbf{Until} $\breve{d}^{(v)}\leq \tilde{\epsilon}$.\\
		11: \textbf{Output:} the optimal passive beamforming associated\\
		\qquad with the time delays, i.e., $\boldsymbol{\vartheta}^\mathrm{opt}$. \\
		\bottomrule[2pt]
	\end{tabular}
\end{center}

In the outer loop iteration, the Lagrange multipliers in $\boldsymbol{\lambda}$ and the penalty coefficient $\breve{\rho}$ will be updated according to difference value of $\breve{d}=\|\boldsymbol{\vartheta}-\widehat{\boldsymbol{\vartheta}}\|_2$. Specifically, in the $(v+1)$-th outer loop iteration, when $\breve{d}^{(v+1)}\leq \epsilon^{(v+1)}=0.9\breve{d}^{(v)}$,  the Lagrange multipliers in $\boldsymbol{\lambda}$ will be updated with the rule of
$
	\boldsymbol{\lambda}\leftarrow \boldsymbol{\lambda}+\frac{\boldsymbol{\vartheta}-\widehat{\boldsymbol{\vartheta}}}{\breve{\rho}}.
$
Otherwise, we will update the penalty coefficient using rule:
$
	\breve{\rho}\leftarrow \breve{\xi}\breve{\rho},
$
 where $\breve{\xi}\in(0,1)$ denotes the scaling factor. Specifically, Algorithm 1 summarizes the detailed procedures for solving the optimization problem \eqref{eq_time_delay_trans_mmse}.
\subsubsection{Modulation Frequency Design} In this subsection, we turn to design the modulation frequencies with the achieved passive beamforming $\boldsymbol{\vartheta}$. Under this setting, the original problem for maximizing $R_\mathrm{b}$ is equivalent to maximizing $\tilde{g}(\mathbf{f})=\left|\mathbf{h}_\mathrm{rb}^H\boldsymbol{\Theta}\boldsymbol{\Theta}_0\mathbf{h}_\mathrm{ar}\right|^2$, then the corresponding subproblem related to modulation frequency design can be formulated as
\begin{subequations}\label{eq_modulation_fre}
	\begin{align}
		&\max _{\mathbf{f}} ~\tilde{g}(\mathbf{f})=\left|\mathbf{h}_\mathrm{rb}^H\widehat{\mathbf{h}}\right|^2,\notag \\
		&~\text { s.t. }  \Delta f_{\min}\leq\Delta f_l\leq \Delta f_{\max},~ l\in\mathcal{L}, \label{eq_modulation_fre_1}\\
		&\qquad \widehat{g}_k(\mathbf{f})=\left|(\mathbf{h}^\mathrm{LoS}_{\mathrm{rw}k})^H\widehat{\mathbf{h}}\right|^2\leq \iota_k, ~k\in\mathcal{K}\label{eq_modulation_fre_2}.
	\end{align}
\end{subequations}
where $\widehat{\mathbf{h}}=\boldsymbol{\Theta}\boldsymbol{\Theta}_0\mathbf{h}_\mathrm{ar}\triangleq[A_1e^{j\varphi_1}, \cdots, A_Le^{j\varphi_L}]$, and the covert constraint \eqref{eq_time_delay_trans_2} can be equivalently transformed into \eqref{eq_modulation_fre_2}. Note that, solving this subproblem is nontrivial since both the LoS components of the legitimate channel $\mathbf{h}_\mathrm{rb}^\mathrm{LoS}$ and the warden channels $\mathbf{h}_{\mathrm{rw}k}^\mathrm{LoS}$ are related to  the modulation frequency variables in $\mathbf{f}$, where strong couplings exist. To handle this issue, the second Taylor expansion theorem \cite{sun2016majorization} is utilized to  construct the concave lower bound of $\tilde{g}$ and convex upper bound of $\widehat{g}_k, k\in\mathcal{K}$. In particular, in the $(p+1)$-th iteration, the concave lower bound of $\tilde{g}$ can be given by
\begin{align}\label{eq_tilde_g_low}
	\tilde{g}(\mathbf{f})\geq& \tilde{g}(\mathbf{f}^{(p)})+\nabla \tilde{g}^T(\mathbf{f}^{(p)})(\mathbf{f}-\mathbf{f}^{(p)})-\notag\\
	&\frac{\nu}{2}(\mathbf{f}-\mathbf{f}^{(p)})^T(\mathbf{f}-\mathbf{f}^{(p)})=\check{g}(\mathbf{f}, \mathbf{f}^{(p)}),
\end{align}
where $\nabla\tilde{g}(\mathbf{f})$ can be derived according to
\begin{itemize}
	\item $\mathcal{I}_l\left(\nabla\tilde{g}(\mathbf{f})\right)=2\operatorname{Re}\left(\frac{\partial \mathbf{h}^H_\mathrm{rb}}{\partial \Delta f_l}\widehat{\mathbf{h}}\widehat{\mathbf{h}}^H\mathbf{h}_\mathrm{rb}\right),$
	\item $
		\mathcal{I}_z\left(\frac{\partial \mathbf{h}_\mathrm{rb}}{\partial \Delta f_l}\right)=\begin{cases}
		j2\pi g\frac{d_l^\mathrm{rb}}{c}\beta_1\rho(d^\mathrm{rb})\times\\
		e^{j2\pi\left(g\Delta f_l\frac{d_l^\mathrm{rb}}{c}+\frac{\Upsilon_l^\mathrm{rb}f_\mathrm{c}}{c}\right)},& z=l,\\
		0,& otherwise.
	\end{cases}$
\end{itemize}
And $\nu$ is a positive number satisfying $\nu\mathbf{I}_{L\times L}\succeq\nabla^2\tilde{g}(\mathbf{f})$\footnote{To ensure the condition $\nu\mathbf{I}_{L\times L}\succeq\nabla^2\tilde{g}(\mathbf{f})$ is consistently satisfied, $\nu$ is typically selected as the largest singular value of the Hessian matrix $\nabla^2\tilde{g}(\mathbf{f})$.} with $\nabla^2\tilde{g}(\mathbf{f})$ denoting the Hessian matrix of $\tilde{g}(\mathbf{f})$ given below
\begin{align}
	\nabla^2\tilde{g}(\mathbf{f})=
		\begin{bmatrix}
			\frac{\partial^2\tilde{g}(\mathbf{f})}{\partial^2 \Delta f_1} & \frac{\partial^2\tilde{g}(\mathbf{f})}{\partial \Delta f_1\partial \Delta f_2} &\cdots &\frac{\partial^2\tilde{g}(\mathbf{f})}{\partial \Delta f_1\partial \Delta f_L} \\
			\frac{\partial^2\tilde{g}(\mathbf{f})}{\partial \Delta f_2\partial \Delta f_1} & \frac{\partial^2\tilde{g}(\mathbf{f})}{\partial^2 \Delta f_2} &  \cdots & \frac{\partial^2\tilde{g}(\mathbf{f})}{\partial \Delta f_2\partial \Delta f_L} \\
			\vdots & \vdots  & \ddots & \vdots\\
			\frac{\partial^2\tilde{g}(\mathbf{f})}{\partial \Delta f_L\partial \Delta f_1} & \frac{\partial^2\tilde{g}(\mathbf{f})}{\partial \Delta f_L\partial \Delta f_2} & \cdots & \frac{\partial^2\tilde{g}(\mathbf{f})}{\partial^2 \Delta f_L}
		\end{bmatrix},
\end{align}
where the expressions of $\frac{\partial^2\tilde{g}(\mathbf{f})}{\partial^2 \Delta f_l}, ~l\in\mathcal{L}$, and $\frac{\partial^2\tilde{g}(\mathbf{f})}{\partial \Delta f_m\partial \Delta f_n}, ~m\neq n, ~m, n \in\mathcal{L}$, are respectively presented in \eqref{eq_tilde_g_ll} and \eqref{eq_tilde_g_mn}, shown at the top of the next page..
\begin{figure*}[t]
	\begin{align}
		&\frac{\partial^2\tilde{g}(\mathbf{f})}{\partial^2 \Delta f_l}=2\operatorname{Re}\left(-j2\pi g\frac{d_l^\mathrm{rb}}{c}\frac{\partial \mathbf{h}^H_\mathrm{rb}}{\partial \Delta f_l}\widehat{\mathbf{h}}\widehat{\mathbf{h}}^H\mathbf{h}_\mathrm{rb}+j2\pi g\frac{d_l^\mathrm{rb}}{c}\beta_1\rho(d^\mathrm{rb})A_le^{-j\varphi_l}	e^{j2\pi\left(g\Delta f_l\frac{d_l^\mathrm{rb}}{c}+\frac{\Upsilon_l^\mathrm{rb}f_\mathrm{c}}{c}\right)}\frac{\partial \mathbf{h}^H_\mathrm{rb}}{\partial \Delta f_l}\widehat{\mathbf{h}}\right), ~l\in\mathcal{L},\label{eq_tilde_g_ll}\\
		&\frac{\partial^2\tilde{g}(\mathbf{f})}{\partial \Delta f_m\partial \Delta f_n}=2\operatorname{Re}\left(j2\pi g\frac{d_n^\mathrm{rb}}{c}A_n\beta_1\rho(d^\mathrm{rb})e^{-j\varphi_n}	e^{j2\pi\left(g\Delta f_n\frac{d_n^\mathrm{rb}}{c}+\frac{\Upsilon_n^\mathrm{rb}f_\mathrm{c}}{c}\right)}\frac{\partial \mathbf{h}^H_\mathrm{rb}}{\partial \Delta f_m}\widehat{\mathbf{h}}\right)
		, ~m\neq n, ~m, n \in\mathcal{L}\label{eq_tilde_g_mn}.
	\end{align}
		\hrule
\end{figure*}

Similarly, the upper-bound of $\widehat{g}_k(\mathbf{f})$ in $(p+1)$-th iteration can be expressed as
\begin{align}
	\widehat{g}_k(\mathbf{f})\leq&\widehat{g}_k(\mathbf{f}^{(p)})+\nabla \widehat{g}_k^T(\mathbf{f}^{(p)})(\mathbf{f}-\mathbf{f}^{(p)})+\notag\\
	&\frac{\widehat{\nu}_k}{2}(\mathbf{f}-\mathbf{f}^{(p)})^T(\mathbf{f}-\mathbf{f}^{(p)})=\overline{g}(\mathbf{f}, \mathbf{f}^{(p)}),
\end{align}
where $\nabla\widehat{g}_k(\mathbf{f})$ can be derived according to
\begin{itemize}
	\item $\mathcal{I}_l\left(\nabla\widehat{g}_k(\mathbf{f})\right)=2\operatorname{Re}\left(\frac{\partial (\mathbf{h}^\mathrm{LoS}_{\mathrm{rw}k})^H}{\partial \Delta f_l}\widehat{\mathbf{h}}\widehat{\mathbf{h}}^H\mathbf{h}^\mathrm{LoS}_{\mathrm{rw}k}\right),$
\item $	\mathcal{I}_z\left(\frac{\partial \mathbf{h}^\mathrm{LoS}_{\mathrm{rw}k}}{\partial \Delta f_l}\right)=\begin{cases}
				e^{j2\pi\left(g\Delta f_l\frac{d_l^{\mathrm{rw}k}}{c}+\frac{\Upsilon_l^{\mathrm{rw}k}f_\mathrm{c}}{c}\right)}\times\\
				j2\pi g\frac{d_l^{\mathrm{rw}k}}{c},& z=l,\\
		0,& otherwise.\\
	\end{cases}$
\end{itemize}
Additionally, $\nu_k$ needs to satisfy: $\nu_k\mathbf{I}_{L\times L}\succeq \nabla^2\widehat{g}_k(\mathbf{f})$. And  $\nabla^2\widehat{g}_k(\mathbf{f})$ can be derived based on expressions \eqref{eq_widehat_g_ll} and \eqref{eq_widehat_g_mn}.
\begin{figure*}[t]
	\begin{align}
		&\hspace{-2mm}\frac{\partial^2\widehat{g}_k(\mathbf{f})}{\partial^2 \Delta f_l}=2\operatorname{Re}\left(-j2\pi g\frac{d_l^{\mathrm{rw}k}}{c}\frac{\partial (\mathbf{h}^\mathrm{LoS}_{\mathrm{rw}k})^H}{\partial \Delta f_l}\widehat{\mathbf{h}}\widehat{\mathbf{h}}^H\mathbf{h}^\mathrm{LoS}_{\mathrm{rw}k}+j2\pi g\frac{d_l^{\mathrm{rw}k}}{c}A_le^{-j\varphi_l}e^{j2\pi\left(g\Delta f_l\frac{d_l^{\mathrm{rw}k}}{c}+\frac{\Upsilon_l^{\mathrm{rw}k}f_\mathrm{c}}{c}\right)}\frac{\partial (\mathbf{h}^\mathrm{LoS}_{\mathrm{rw}k})^H}{\partial \Delta f_l}\widehat{\mathbf{h}}\right), ~l\in\mathcal{L},\label{eq_widehat_g_ll}\\
		&\hspace{-2mm}\frac{\partial^2\widehat{g}_k(\mathbf{f})}{\partial \Delta f_m\partial \Delta f_n}=2\operatorname{Re}\left(j2\pi g\frac{d_n^{\mathrm{rw}k}}{c}A_ne^{-j\varphi_n}	e^{j2\pi\left(g\Delta f_n\frac{d_n^{\mathrm{rw}k}}{c}+\frac{\Upsilon_n^{\mathrm{rw}k}f_\mathrm{c}}{c}\right)}\frac{\partial (\mathbf{h}^\mathrm{LoS}_{\mathrm{rw}k})^H}{\partial \Delta f_m}\widehat{\mathbf{h}}\right)
		, ~m\neq n, ~m, n \in\mathcal{L}\label{eq_widehat_g_mn}.
	\end{align}
	\hrule
\end{figure*}
\begin{figure*}[ht]
	\centering
	\includegraphics[scale=0.41]{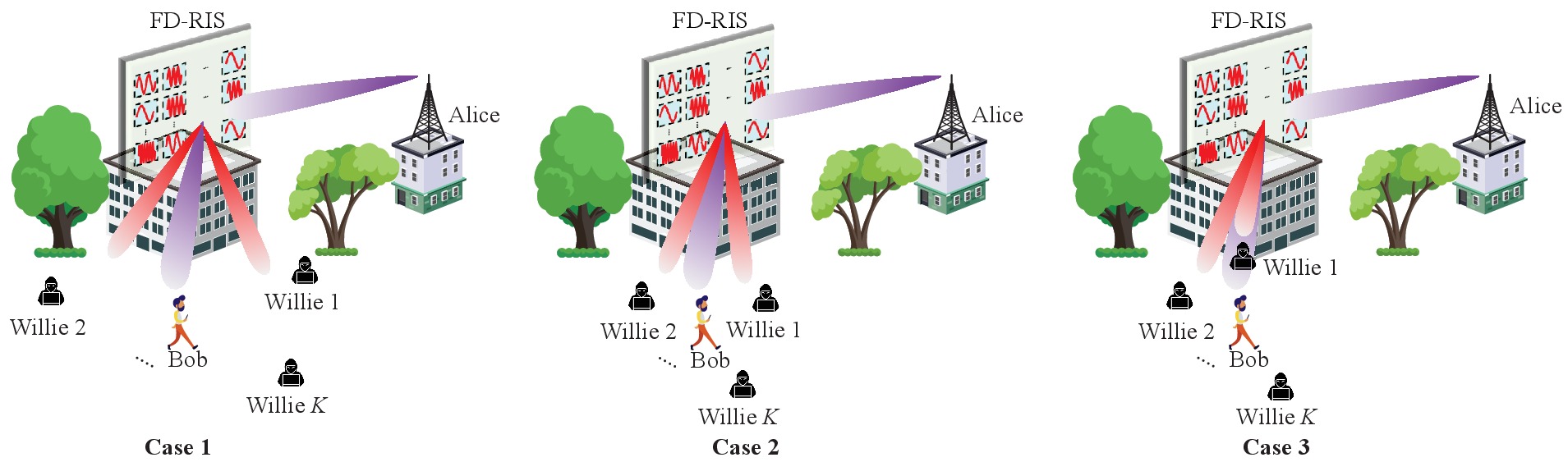}\\
	\caption{FD-RIS-supported three covert communication scenarios.}\label{fig:three_scenarios}
\end{figure*}
 Therefore, the subproblem \eqref{eq_modulation_fre} can be converted into a convex optimization problem in the $(p+1)$-th iteration, given by
 \begin{subequations}\label{eq_modulation_fre_trans}
 	\begin{align}
 		&\max _{\mathbf{f}} ~\check{g}(\mathbf{f}, \mathbf{f}^{(p)}),\notag \\
 		&~\text { s.t. }  \Delta f_{\min}\leq\Delta f_l\leq \Delta f_{\max},~ l\in\mathcal{L}, \label{eq_modulation_fre_trans_1}\\
 		&\qquad \overline{g}_k(\mathbf{f}, \mathbf{f}^{(p)})\leq \iota_k, ~k\in\mathcal{K}\label{eq_modulation_fre_trans_2}.
 	\end{align}
 \end{subequations}
This convex problem can also be effectively addressed using the CVX solver.

\subsection{Analysis on Computing Complexity and Convergence}
The solving procedures for the original optimization problem \eqref{eq_ori_opt} is summarized in Algorithm 2, which leverages the MMSE method, the PDD algorithm framework and the SCA technique. Algorithm 2 iteratively updates the time delay and modulation frequency variables until the difference in the objective function values between successive iterations falls below a predefined threshold $\breve{\epsilon}$.
\begin{center}
	\begin{tabular}{p{8.5cm}}
		\toprule[2pt]
		\textbf{Algorithm 2:}  Proposed Iterative Algorithm for Solving the Original Optimization Problem \eqref{eq_ori_opt}   \\
		\midrule[1pt]
		1: Initialize  $\big(\boldsymbol{\vartheta}^{(0)}, \mathbf{f}^{(0)}\big)$, the tolerance accuracy $\breve{\epsilon}$; Set iter-\\
		\quad ation index $p=$0.\\
		2: \textbf{Repeat} \\
		3:\quad Transform the subproblem \eqref{eq_time_delay_trans} as \eqref{eq_time_delay_trans_mmse} utilizing the\\
		\qquad MMSE method, given the values of $\big(\boldsymbol{\vartheta}^{(p)}, \mathbf{f}^{(p)}\big)$.
		\\
		4:\quad Algorithm 1 is utilized to solve optimization problem\\
		\qquad\eqref{eq_time_delay_trans_mmse} and obtain $\boldsymbol{\vartheta}^{(p+1)}$ with given $\mathbf{f}^{(p)}$.\\
		5:\quad Solve subproblem \eqref{eq_modulation_fre_trans} with obtained $\boldsymbol{\vartheta}^{(p+1)}$ and  achi-\\ \qquad eve $\mathbf{f}^{(p+1)}$. Calculate $R_\mathrm{b}^{(p+1)}$ using $\big(\boldsymbol{\vartheta}^{(p+1)}, \mathbf{f}^{(p+1)}\big)$. \\ \qquad Let
		$p\leftarrow p+1$. \\
		6: \textbf{Until} $R_\mathrm{b}^{(p+1)}-R_\mathrm{b}^{(p)}\leq \breve{\epsilon}$.\\
		7: \textbf{Output:} the optimal time delays $\boldsymbol{\kappa}^\mathrm{opt}$ and modulation\\
		\quad frequencies $\mathbf{f}^\mathrm{opt}$. \\
		\bottomrule[2pt]
	\end{tabular}
\end{center}

Next, we will analyze the computational complexity of the proposed algorithm. Specifically, the computational burden of the iterative algorithm primarily arises from the process of addressing time-delay subproblem \eqref{eq_time_delay_trans_mmse} and modulation frequency subproblem \eqref{eq_modulation_fre}. Regarding the subproblem \eqref{eq_time_delay_trans_mmse}, the PDD algorithm framework is leveraged to transform \eqref{eq_time_delay_trans_mmse} as two quadratically constrained quadratic programming (QCQP) problems, i.e., \eqref{eq_time_delay_trans_PDD_al1} and \eqref{eq_time_delay_trans_PDD_al2}. Actually, the computational complexity associated with solving \eqref{eq_time_delay_trans_PDD_al1} constitutes the dominant part in addressing \eqref{eq_time_delay_trans_mmse}. The computational complexity can be quantified as $\mathcal{C}_1 = \mathcal{O}(I_1 L^{3.5})$, where $I_1$ denotes the total number of iterations in \textbf{Algorithm 1}. This is substantially lower than the complexity of the semidefinite relaxation (SDR)-based method, which scales as $\mathcal{O}(L^7)$, particularly when $L$ is large. For the second subproblem, with the help of second-order Taylor expansion technique, modulation frequency subproblem \eqref{eq_modulation_fre} is also approximately converted as a convex QCQP problem \eqref{eq_modulation_fre_trans}. The computation complexity of solving this problem can be calculated as $\mathcal{C}_2=\mathcal{O}(L^{3.5})$. On the basis of analysis above, the total computing complexity of \textbf{Algorithm 2} is $\mathcal{C}_\mathrm{tol}=I_2\mathcal{C}_1+I_2\mathcal{C}_2$, where $I_2$ denotes the total number of iteration for  \textbf{Algorithm 2}, respectively.

The alternative strategy is adopted to decompose the original optimization problem into two subproblems, respectively for the time-delay and modulation frequency design. These subproblems are then solved iteratively, where each subproblem is addressed using the solution obtained from the previous iteration, which ensures that the objective function value is monotonically non-decreasing at each iteration, thereby guaranteeing convergence to at least a local optimum.
The convergence of the proposed algorithm will be further validated through numerical simulations. 

\section{Numerical Simulations}\label{sec:S5}
In this section, three CC scenarios featuring various spatial correlations between Bob and wardens are investigated via numerical simulations to evaluate the potential of FD-RIS in enhancing the covert performance of wireless systems. The specific details of these scenarios are provided as follows:

 \textbf{Case 1:} In this scenario, the spatial correlation between Bob and wardens is assumed to be negligible, which is a common assumption in conventional RIS-assisted CC schemes. This setting corresponds to Case 1 illustrated in the Fig.~\ref{fig:three_scenarios}.

 \textbf{Case 2:} Scenario 2, as presented in Case 2 of the Fig.~\ref{fig:three_scenarios}, considers a high spatial correlation between Bob and wardens (for example, Willie $K$), which poses challenges for conventional RIS-assisted systems in maintaining CCs.
\begin{table}[h!]
	\renewcommand\arraystretch{1.5}
	\centering
	\caption{Parameters Setting}\label{tab:table1}
	\resizebox{.8\columnwidth}{!}{
		\begin{tabular}{M{3.8cm}|M{4cm}}
			\hline\hline
			\textbf{\normalsize{Parameters}} & \textbf{\normalsize{Symbol and Value}}\\
			\hline\hline
			\normalsize{Alice's position}&\normalsize{$(\theta^\mathrm{ar}, \phi^\mathrm{ar}, d^\mathrm{ar})=(70^\circ, 10^\circ, 70$ m$)$ }   \\
			\hline
			\normalsize{Bob's position}& \normalsize{$(\theta^\mathrm{rb}, \phi^\mathrm{rb}, d^\mathrm{rb})=(120^\circ, 30^\circ, 20$ m$)$}  \\
			\hline
			\normalsize{Carrier frequency}& \normalsize{$f_\mathrm{c}=28$ GHz} \\
			\hline
			\normalsize{Large-scale path loss}& \normalsize{$\rho^2(\widehat{d})=-45-20\log_{10}^{\widehat{d}}$ dB}\\
			\hline
			\normalsize{Lower and upper bound of modulation frequency}& \normalsize{$\Delta f_{\min}=10$ MHz, $\Delta f_{\max}=30$ MHz}\\
			\hline
			\normalsize{Noise power}& \normalsize{$\sigma_\mathrm{b}^2=\widehat{\sigma}_{\mathrm{w}k}^2=-110$ dBm}   \\
			\hline
			\normalsize{\textcolor{blue}{Parameters associated with Algorithm 1}}&
			\normalsize{\textcolor{blue}{$\boldsymbol{\lambda}^{(0)}=\mathbf{0}$, $\breve{\rho}=100$, $\xi=0.5$}}\\
			\hline
			\normalsize{Penalty exponent}& \normalsize{$\psi=100$}  \\
			\hline
			\normalsize{Power budget}& \normalsize{$P_\mathrm{t}=15$ dBm}\\
			\hline
			\normalsize{Rician factor}& \normalsize{$\beta=15$ dB}\\
			\hline
			\normalsize{The number of wardens}& \normalsize{ $K=4$}\\
			\hline
			\normalsize{\textcolor{blue}{Tolerance accuracy}}& \normalsize{ \textcolor{blue}{$\tilde{\epsilon}=\breve{\epsilon}=10^{-3}$}}\\
			\hline
		\end{tabular}
	}
\end{table}

  \textbf{Case 3:} Building upon Case 2, this scenario explores a more adverse CC condition, in which one of the wardens (Willie 1) not only aligns with Bob in spatial direction but also resides at a shorter distance from the FD-RIS, thereby posing a greater threat to covert transmission, as shown in Case 3 of the Fig.~\ref{fig:three_scenarios}. Note that in this extreme scenario, traditional RIS-assisted systems are incapable of preventing eavesdropping, thereby compromising the covert nature of the transmission.

Additionally, the following two baseline schemes are compared to validate the effectiveness of the proposed FD-RIS-aided CC scheme and the iterative algorithm: \textbf{1) RIS-assisted scheme:} In this scheme, the traditional RIS replaces the FD-RIS to implement covert transmission between Alice and Bob under different scenarios. \textbf{2) SDR scheme:} The SDR technique is employed as a substitute for the PDD method to solve the subproblem \eqref{eq_time_delay_trans_mmse}. Here, Table \ref{tab:table1} presents the system parameters utilized in numerical simulations. In addition,  Willies' positions in Case 1 are set as: $(\theta_1^{\mathrm{rw1}}, \theta_1^{\mathrm{rw2}}, \theta_1^{\mathrm{rw3}}, \theta_1^{\mathrm{rw4}})=(85^\circ, 90^\circ, 100^\circ, 145^\circ)$, $(\phi_1^{\mathrm{rw1}},$ $\phi_1^{\mathrm{rw2}}, \phi_1^{\mathrm{rw3}}, \phi_1^{\mathrm{rw4}})=(50^\circ, 15^\circ, 45^\circ, 60^\circ)$
and $(d_1^{\mathrm{rw1}}, d_1^{\mathrm{rw2}},$ $ d_1^{\mathrm{rw3}},d_1^{\mathrm{rw4}})=(45, 20, 55, 30)$ m.
Willies' positions in Case 2 are set as: $(\theta_2^{\mathrm{rw1}}, \theta_2^{\mathrm{rw2}}, \theta_2^{\mathrm{rw3}}, \theta_2^{\mathrm{rw4}})=(115^\circ, 110^\circ, 110^\circ, 125^\circ)$,
$(\phi_2^{\mathrm{rw1}}, \phi_2^{\mathrm{rw2}}, \phi_2^{\mathrm{rw3}}, \phi_2^{\mathrm{rw4}})=(30^\circ, 25^\circ, 35^\circ, 40^\circ)$
and $(d_2^{\mathrm{rw1}},$ $ d_2^{\mathrm{rw2}},d_2^{\mathrm{rw3}}, d_2^{\mathrm{rw4}})=(45, 20, 55, 30)$ m.
Willies' positions in Case 3 are set as:
$(\theta_3^{\mathrm{rw1}}, \theta_3^{\mathrm{rw2}}, \theta_3^{\mathrm{rw3}}, \theta_3^{\mathrm{rw4}})=(120^\circ, 110^\circ,$ $110^\circ, 125^\circ)$, $(\phi_3^{\mathrm{rw1}}, \phi_3^{\mathrm{rw2}}, \phi_3^{\mathrm{rw3}}, \phi_3^{\mathrm{rw4}})=(30^\circ, 25^\circ, 35^\circ, 40^\circ)$
and $(d_3^{\mathrm{rw1}}, d_3^{\mathrm{rw2}}, d_3^{\mathrm{rw3}}, d_3^{\mathrm{rw4}})=(15, 20, 55, 30)$ m.
 \begin{figure}[ht]
	\centering
\includegraphics[scale=0.31]{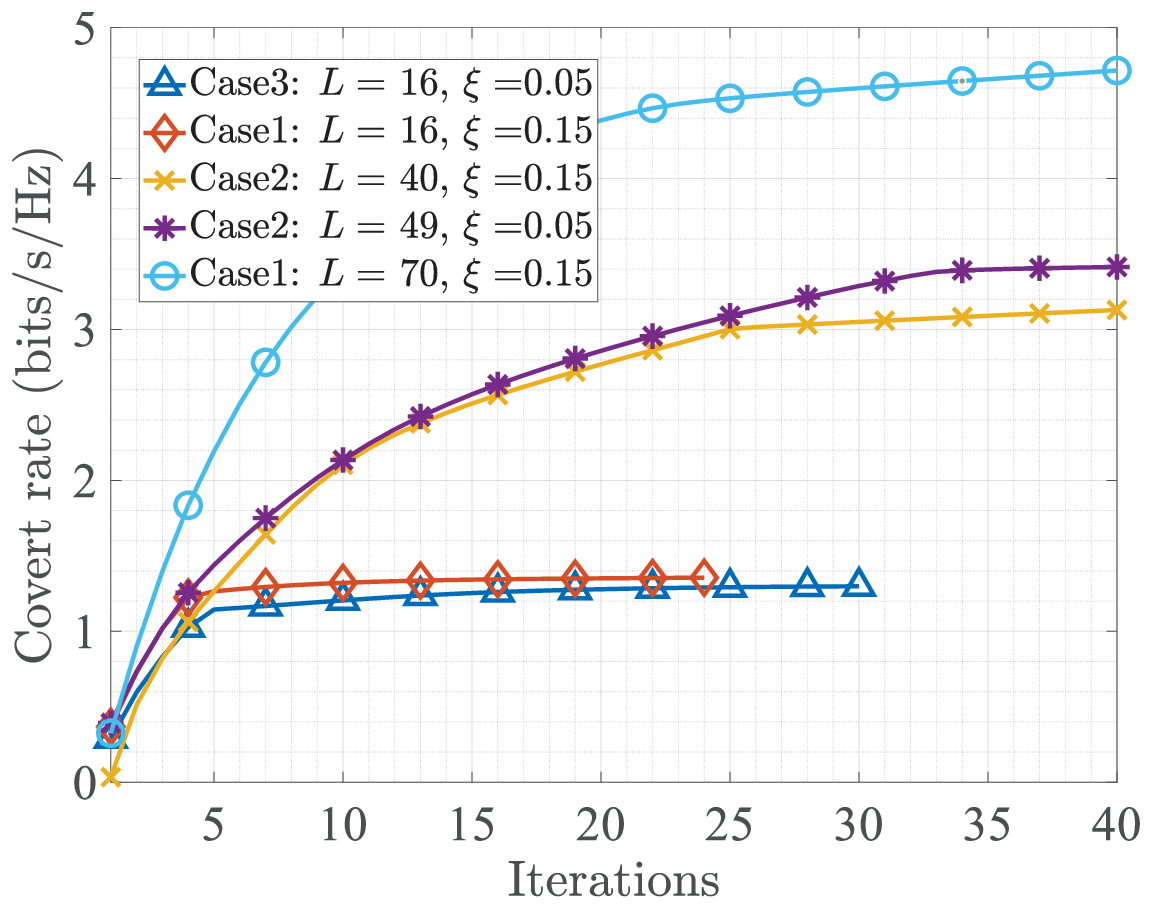}\\
\caption{The convergence analysis of the proposed algorithm.}\label{fig:Iterations}
\end{figure}

\vspace{-2mm}Fig. \ref{fig:Iterations} presents the convergence simulation results of the proposed algorithm taking into account different number of elements and covert requirement. In particular, it is observed that the objective function exhibits a monotonically increasing trend as the number of iterations increases, indicating that the proposed algorithm achieves performance improvement at each iteration. Moreover, as the number of RIS elements increases, the number of iterations required for convergence also grows. This is because the dimension of the optimization variables increases with more elements, leading to a larger search space and thus requiring more iterations to reach a stable solution. These results fully demonstrate that the proposed algorithm possesses good convergence and stability under different system scales.
 \begin{figure}[ht]
	\centering
	\includegraphics[scale=0.26]{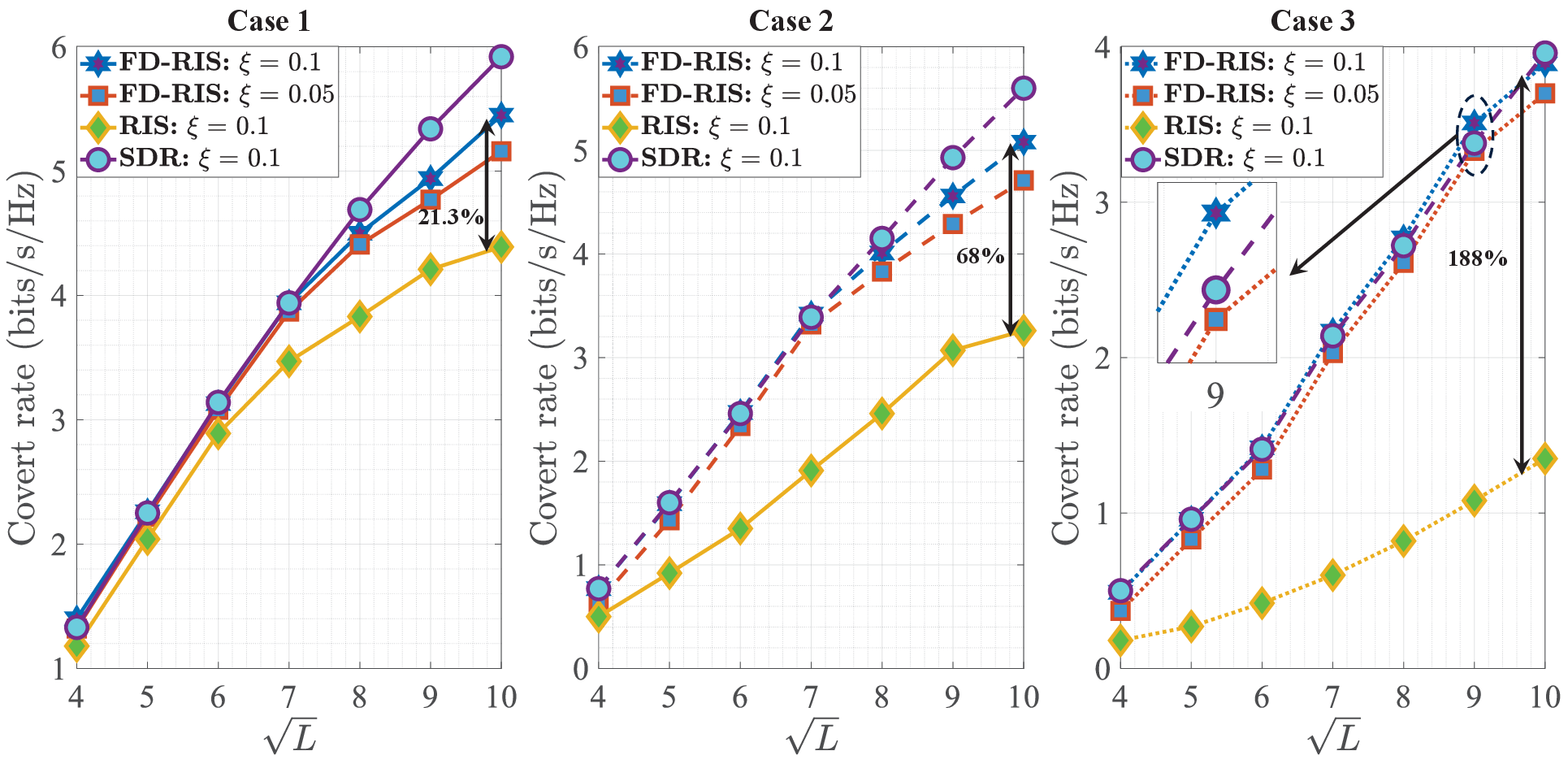}\\
	\caption{Analysis of the covert rate in relation to the number of RIS elements, considering varying covertness demands across three distinct scenarios.}\label{fig:L_vs_CR}
\end{figure}

Fig.~\ref{fig:L_vs_CR} illustrates the variation of covert rate with respect to the number of RIS elements $L$, under different covert constraints $\xi$ and communication scenarios. As $L$ increases, the covert rate consistently improves across all settings. This is because a larger number of RIS elements provides more DoFs, enhancing beamforming precision and energy focusing capability.
 However, the performance gain achieved by the conventional RIS remains limited in all cases. Notably, its covert rate is even lower than that of the FD-RIS operating under stricter covert constraints. As $L$ increases, the performance gap between FD-RIS and conventional RIS schemes becomes more pronounced. For instance, when $L = 100$, the FD-RIS outperforms the conventional RIS by 21.3\%, 68\%, and 188\% across the three evaluated scenarios, respectively, clearly highlighting the significant potentials of FD-RIS in improving CC performance.
 Moreover, increasing spatial correlation between the warden and Bob leads to performance degradation for both schemes, as higher correlation enhances the warden’s detection capability. Nevertheless, the degradation is substantially more severe for the conventional RIS, which only performs angular-domain control and lacks robustness in highly correlated environments. In contrast, the FD-RIS, leveraging joint distance-angle control, remains resilient and maintains high covert performance even when Bob and the warden share similar angular locations.
 \begin{figure}[ht]
	\centering
	\includegraphics[scale=0.27]{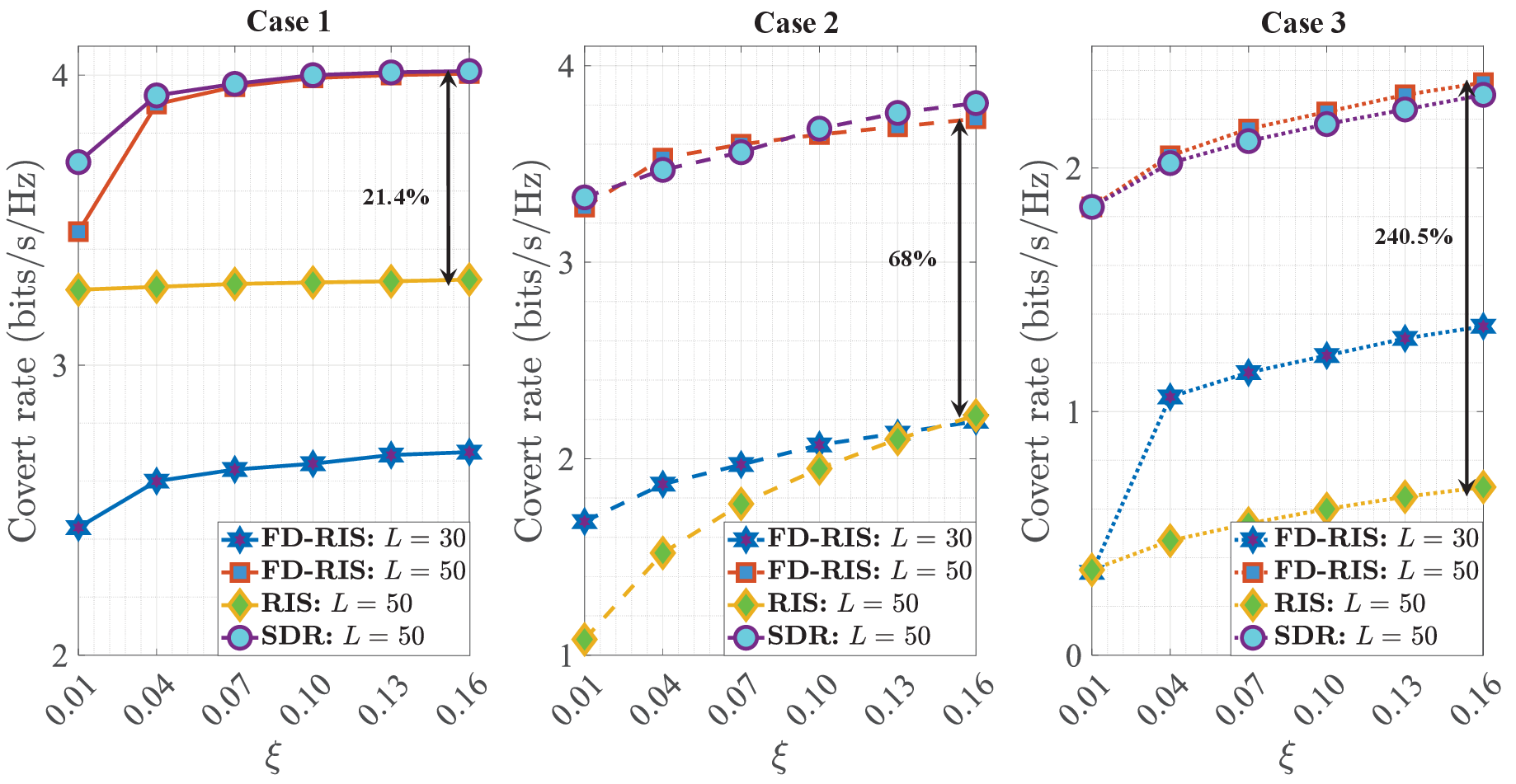}\\
		\caption{Analysis of the covert rate in relation to the covert requirement, considering varying element counts across three distinct scenarios.}\label{fig:xi_vs_CR}
\end{figure}

The impact of the covert requirement $\xi$ on the covert rate is analyzed under varying numbers of RIS elements and different communication scenarios, as shown in Fig. \ref{fig:xi_vs_CR}. The results reveal that as the covert constraint is gradually relaxed, the covert rate steadily increases. However, the rate of growth diminishes due to limitations imposed by system factors such as power budget and channel conditions, which restrict the extent of performance improvement.
Furthermore, although the SDR scheme achieves performance comparable to the proposed algorithm, it suffers from substantially higher computational complexity. This underscores the effectiveness of the proposed algorithm in achieving near-optimal performance with significantly improved computational efficiency.
In addition, FD-RIS demonstrates clear advantages over the conventional RIS in enhancing CC performance, especially under scenarios with high spatial correlation. For instance, when $\xi = 0.16$, the FD-RIS scheme achieves approximately $2.5$ times the covert rate of the conventional RIS. The reason for this substantial improvement lies in the underlying beamforming capabilities: the conventional RIS, which can only manipulate signals in the angular domain, tends to reduce the signal power delivered to Bob in order to satisfy the covert constraint, thereby degrading communication quality. In contrast, the FD-RIS performs joint distance-angle control, enabling it to steer signal energy away from eavesdroppers aligned in direction with the legitimate user, while efficiently focusing more energy toward the intended receiver. This dual-domain control results in significantly better covert performance.
\begin{figure}[ht]
	\centering
	\includegraphics[scale=0.32]{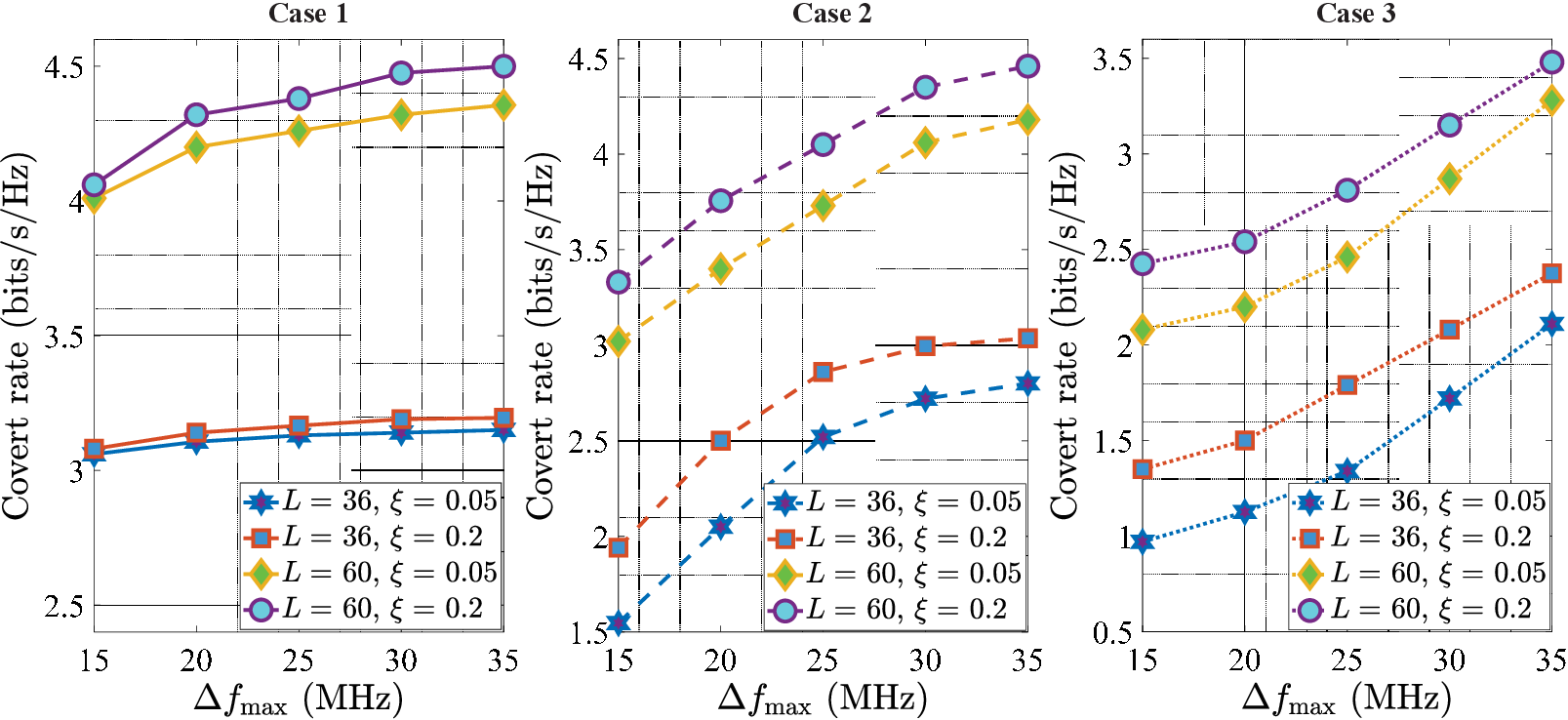}\\
	\caption{Analysis of the covert rate in relation to the upper bound of modulation frequency, considering varying element counts and covert requirements across three distinct scenarios.}\label{fig:f_max_vs_CR}
\end{figure}

\begin{figure*}[ht]
	\centering
	\includegraphics[scale=0.4]{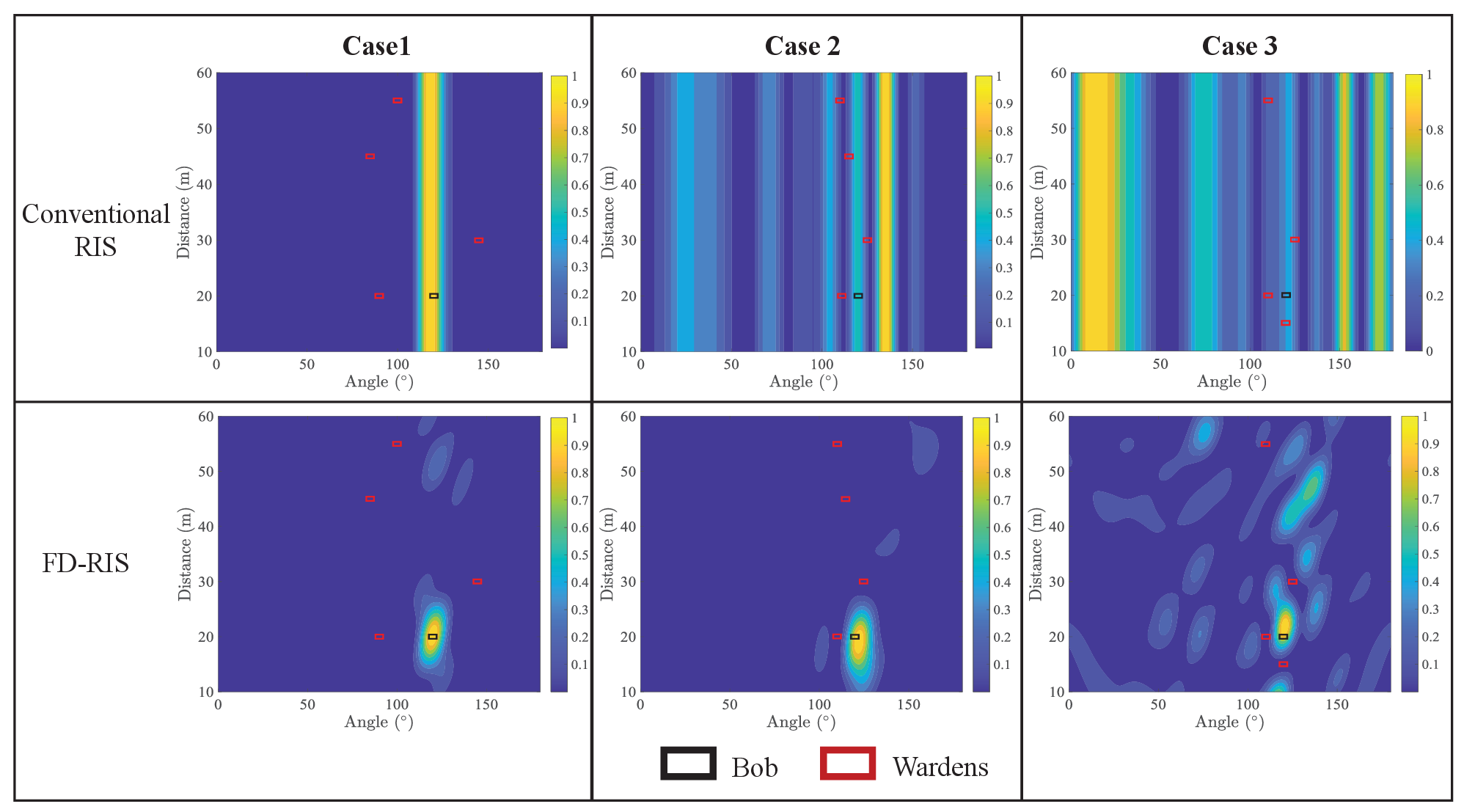}\\
	\caption{Comparative analysis of conventional RIS patterns versus FD-RIS patterns across three distinct scenarios.}\label{fig:pattern_comparison}
\end{figure*}

\textcolor{blue}{To evaluate how modulation frequency affects the covert rate, Fig. \ref{fig:f_max_vs_CR} illustrates the covert rate as a function of the upper bound of the modulation frequency, $\Delta f_{\max}$, under different element counts and covert constraints across three representative scenarios. Overall, across all considered scenarios, the covert rate exhibits an increasing trend as $\Delta f_{\max}$ grows. This is because a larger $\Delta f_{\max}$ provides a wider range of selectable modulation frequencies during the optimization process, enabling the system to identify modulation frequencies that better match the transmission channel, thereby enhancing security performance.
By comparing the trends of the covert rate across the three scenarios, we observe that in Case 1 and Case 2, the growth rate of the covert rate gradually slows down, whereas in Case 3, the covert rate increases in an accelerated manner. This discrepancy stems from the fact that in Case 1 and Case 2, the spatial correlation between wardens and Bob is relatively weak, which limits the effectiveness of distance beamforming in improving the covert rate. Since the modulation frequency is the key parameter that enables distance beamforming, increasing $\Delta f_{\max}$ only provides limited gains in these two scenarios. In contrast, in Case 3, wardens and Bob share the same spatial direction, making distance beamforming critically important for enhancing covert performance. As a result, expanding $\Delta f_{\max}$ becomes highly beneficial, allowing the system to more fully exploit this dimension and achieve significantly higher covert rate improvements.
}

To further illustrate the underlying mechanism by which the FD-RIS enhances CC performance, Fig.~\ref{fig:pattern_comparison} compares the beam-patterns of the conventional RIS and the FD-RIS across three distinct CC scenarios. Note that for clearer visualization, it is assumed that $(\phi_{s}^\mathrm{rw1}, \phi_{s}^\mathrm{rw2}, \phi_{s}^\mathrm{rw3}, \phi_{s}^\mathrm{rw4}) = (30^\circ, 30^\circ, 30^\circ, 30^\circ)$ for all $s \in \{1, 2, 3, 4\}$.
 As observed from the figure, the conventional RIS only possesses beamforming capability in the angular domain. Therefore, when the spatial correlation between the wardens and Bob increases, the conventional RIS can only satisfy the covert constraint by suppressing the signal power directed toward Bob, to reduce the risk of being detected (as evidenced by the decrease in energy along Bob's direction from left to right in the RIS beampattern). This limitation implies that in highly correlated scenarios, the conventional RIS struggles to balance covertness and communication performance, resulting in significantly degraded CC capability.
In contrast, the FD-RIS supports joint beamforming in both distance and angle domains, enabling flexible control over the spatial distribution of signal energy. As a result, it can still effectively focus sufficient energy on Bob while satisfying the covert constraint. This dual-domain control allows FD-RIS to maintain strong covert performance even in extreme cases where the wardens and Bob are aligned in same direction, demonstrating superior robustness and adaptability in challenging spatial environments.

\vspace{-2mm}\section{Conclusion and Prospect}\label{sec:S6}
This paper initially proposes the novel FD-RIS-assisted CC scheme. In particular, the signal processing model of the FD-RIS, incorporating effective control of harmonic \mbox{signals} and time-delay techniques, is first derived. Furthermore, the distance-angle beamforming capability of the FD-RIS is demonstrated based on the beampattern. Subsequently, to evaluate the potentials of the FD-RIS, \textcolor{blue}{we integrate the FD-RIS into the CC system with multiple wardens and derive the analytical covert constrain using the logarithmic moment generating function under an extreme wiretapping scenario.
In this case, the time-delay and modulation frequency variables are jointly optimized to maximize the covert user Bob's achievable rate taking account of covert constrains. To address the challenge of strong coupling between variables, the alternative strategy is first utilized to divide the optimization problem into two subproblems, i.e., active and passive beamforming subproblems. The MMSE method, the PDD algorithm framework and the SCA technique are leveraged to solve these two subproblems.} To validate and quantify the effectiveness of the proposed scheme, extensive simulations are conducted. The results demonstrate that, empowered by joint distance-angle beamforming, the FD-RIS significantly outperforms the conventional RIS in enhancing covert performance, particularly under extreme scenarios with high spatial correlation between Bob and wardens. \textcolor{blue}{Note that the proposed FD-RIS-aided CC transmission scheme offers a significant advantage in mitigating the effects caused by security blind zone inherent in traditional RIS-aided systems, while also providing a solid theoretical foundation for the practical application of FD-RIS in wireless communication systems.}

\textcolor{blue}{This study focuses on a scenario in which Alice possesses perfect position information of potential wardens, who are assumed to be passive and non-colluding.
	In future work, it would be valuable to investigate more practical CC scenarios, such as colluding wiretapping cases and hybrid wiretapping environments involving both active and passive wardens, and to extend the analysis to these settings. Moreover, integrating FD-RIS into physical layer security (PLS) frameworks represents another promising research direction. These extensions are expected to further enhance the applicability and robustness of FD-RIS-assisted systems in practical and dynamically changing communication environments.}

\appendices
\section{Proof of Theorem \ref{th1}}\label{appedix 1}
According to the DEP of Willie $k$ in \eqref{eq_DEP}, $P_{\mathrm{e}, k}$ is a segment function considering two ranges. Thus, we will derive the optimal detection threshold from these two range of $\tau_{{\mathrm{dt}}, k}$.

\textit{1):} when $\tau_{{\mathrm{dt}}, k}\geq \frac{\widehat{\sigma}^2_{\mathrm{w}k}}{\varsigma}+\omega_k$, $P_{\mathrm{e}, k}= 1-\frac{\left(\ln\tau_{{\mathrm{dt}}, k}-\ln(\tau_{{\mathrm{dt}}, k}-\omega_k )\right)}{2\ln\varsigma}$. Hence, the first-order partial derivative of $P_{\mathrm{e}, k}$ w.r.t. $\tau_{{\mathrm{dt}}, k}$ can be expressed as
\vspace{-2mm}\begin{align}
	\frac{\partial P_{\mathrm{e}, k}}{\partial \tau_{{\mathrm{dt}}, k}}=\frac{\omega_k}{2\ln\varsigma(\tau_{{\mathrm{dt}}, k}-\omega_k)\tau_{{\mathrm{dt}}, k}}.
\end{align}
Note that $\frac{\partial P_{\mathrm{e}, k}}{\partial \tau_{{\mathrm{dt}}, k}}\geq 0$ always holds, indicating that $P_{\mathrm{e}, k}$ monotonically increases versus $\tau_{{\mathrm{dt}}, k}$, thus the minimum $P_{\mathrm{e}, k}$ is achieved at the optimal detection threshold $\tau_{{\mathrm{dt}}, k}^\mathrm{opt}= \frac{\widehat{\sigma}^2_{\mathrm{w}k}}{\varsigma}+\omega_k$.

\textit{2):} when $\tau_{{\mathrm{dt}}, k}< \frac{\widehat{\sigma}^2_{\mathrm{w}k}}{\varsigma}+\omega_k$, $P_{\mathrm{e}, k}= 1-\frac{(\ln\tau_{{\mathrm{dt}}, k}-\ln( \frac{\widehat{\sigma}^2_{\mathrm{w}k}}{\varsigma}))}{2\ln\varsigma}$. Hence, the first-order partial derivative of $P_{\mathrm{e}, k}$ w.r.t. $\tau_{{\mathrm{dt}}, k}$ can be expressed as
$
	\frac{\partial P_{\mathrm{e}, k}}{\partial \tau_{{\mathrm{dt}}, k}}=-\frac{1}{2\ln\varsigma\tau_{{\mathrm{dt}}, k}}<0,
$
which indicates $P_{\mathrm{e}, k}$ is a monotonically decreasing function w.r.t. $\tau_{{\mathrm{dt}}, k}$ in the considering range.

According to the above discussion, and considering both the continuity of $P_{\mathrm{e},k}$ at the point $\tau_{\mathrm{dt},k} = \frac{\widehat{\sigma}^2_{\mathrm{w}k}}{\varsigma} + \omega_k$ and the feasible range of $\tau_{\mathrm{dt},k}$, the optimal detection threshold can be derived as $
	\tau_{{\mathrm{dt}}, k}^\mathrm{opt}=\min\left\{\omega_k+\frac{\widehat{\sigma}^2_{\mathrm{w}k}}{\varsigma},~ \varsigma\widehat{\sigma}^2_{\mathrm{w}k}\right\}.
$

\section{Proof of Theorem \ref{th2}}\label{appedix 2}
In this section, the analytical expression of $\tilde{\omega}_k$ will be derived. Specifically, let $X_k = \sqrt{P_\mathrm{t}} \mathbf{h}_{\mathrm{rw}k}^H \boldsymbol{\Theta} \boldsymbol{\Theta}_0 \mathbf{h}_\mathrm{ar}$. Based on the distribution of $\mathbf{h}^{\mathrm{NLoS}}_{\mathrm{rw}k}$, we have
$X_k \sim \mathcal{CN}(\mu_k, \tilde{\sigma}_k^2)$,
where
$\mu_k = \sqrt{P_\mathrm{t}} \rho(d^{\mathrm{rw}k}) \beta_1 (\mathbf{h}^{\mathrm{LoS}}_{\mathrm{rw}k})^H \boldsymbol{\Theta} \boldsymbol{\Theta}_0 \mathbf{h}_\mathrm{ar}$
and
$\tilde{\sigma}_k^2 = P_\mathrm{t} \rho^2(d^{\mathrm{rw}k}) \left| \beta_2 \boldsymbol{\Theta} \boldsymbol{\Theta}_0 \mathbf{h}_\mathrm{ar} \right|^2$.
Let $X_k = \frac{\tilde{\sigma}k}{\sqrt{2}} X_{1k} + i \frac{\tilde{\sigma}k}{\sqrt{2}} X_{2k}$, where $X_{1k}$ and $X_{2k}$ are independent standard real Gaussian random variables. Then we have:
$X_{1k}^2 \sim \chi_1^2\left(\frac{2\operatorname{Re}(\mu_k)^2}{\tilde{\sigma}_k^2}\right)$ and
$X_{2k}^2 \sim \chi_1^2\left(\frac{2\operatorname{Im}(\mu_k)^2}{\tilde{\sigma}_k^2}\right)$.

Therefore, the random variable
$G = X_{1k}^2 + X_{2k}^2$
follows a non-central chi-square distribution:
$G \sim \chi_2^2\left(\frac{2|\mu_k|^2}{\tilde{\sigma}_k^2}\right)$.
The moment generating function of $G$ is given by:
\vspace{-2mm}
\begin{align}
	M_G(s)=\mathbb{E}(e^{sG})=\frac{e^{\frac{2|\mu_k|^2s}{\tilde{\sigma}^2_k(1-2s)}}}{1-2s},~ s<\frac{1}{2}.
\end{align}
Let $s=\frac{\tilde{\sigma}^2_k}{2}\psi$, we have
$
	\mathbb{E}(e^{\psi\omega_k})=\frac{e^{\frac{\psi|\mu_k|^2}{(1-\psi\tilde{\sigma}^2_k)}}}{1-\psi\tilde{\sigma}^2_k}$ with $\tilde{\sigma}^2_k\psi<1
$.
Hence, we can derive that
\begin{align}
	\tilde{\omega}_k
	=&\frac{\mu_k^2}{1-\psi\tilde{\sigma}^2_k}-\frac{\log(1-\psi\tilde{\sigma}^2_k)}{\psi}.
\end{align}

\ifCLASSOPTIONcaptionsoff
  \newpage
\fi
\bibliographystyle{IEEEtran}
\bibliography{FD-RIS_CC}

\end{document}